\documentclass[a4paper,11pt]{article}

\usepackage{jcappub}

\def\lsim{~\rlap{$<$}{\lower 1.0ex\hbox{$\sim$}}}
\def\bsim{~\rlap{$>$}{\lower 1.0ex\hbox{$\sim$}}}

\def\hmsun{\ {\rm M_\odot/{\it h}}}

\def\la{\langle}
\def\ra{\rangle}

\def\ln{{\rm ln}}

\def\det{{\rm det}}

\def\mathbi#1{\textbf{\em #1}}

\def\ov{\overline}

\def\vx{\mathrm{\bf x}}

\def\vmm{\mathrm{M}}

\def\grad{\mathbi{$\nabla$}}

\def\npk{n_{\rm pk}}
\def\esp{n_{\rm ESP}}
\def\fesp{f_{\rm ESP}}
\def\bnpk{\bar{n}_{\rm pk}}
\def\besp{\bar{n}_{\rm ESP}}
\def\dpk{\delta_{\rm pk}}

\def\xpk{\xi_{\rm pk}^{(2)}}

\def\apjl{Astrophys.\ J.\ Lett.}

\def\aap{Astron.\ Astrophys.}

\def\apjs{Astrophys.\ J. Supp.}

\newcommand{\be}{\begin{equation}}
\newcommand{\ee}{\end{equation}}

\def\mathbi#1{\textbf{\em #1}}

\newcommand{\fnl}{f_{\rm NL}}

\newcommand{\calM}{{\cal M}}

\title{Non-Gaussian bias: insights from discrete density peaks}

\author[a]{Vincent Desjacques,}
\author[b,c,d]{Jinn-Ouk Gong}
\author[a]{and Antonio Riotto}

\affiliation[a]{D\'epartement de Physique Th\'eorique
and 
Center for Astroparticle Physics, 
Universit\'e de Gen\`eve, CH-1211 Gen\`eve, Switzerland}
\affiliation[b]{Theory Division, CERN, CH-1211 Gen\`eve 23, Switzerland}
\affiliation[c]{Asia Pacific Center for Theoretical Physics, Pohang 790-784, Korea}
\affiliation[d]{Department of Physics, Postech, Pohang 790-784, Korea}

\emailAdd{Vincent.Desjacques@unige.ch}
\emailAdd{jinn-ouk.gong@apctp.org}
\emailAdd{Antonio.Riotto@unige.ch}

\abstract{
Corrections induced by primordial non-Gaussianity to the linear halo bias can be 
computed from a peak-background split or the widespread local bias model. However,
numerical simulations clearly support the prediction of the former, in which the
non-Gaussian amplitude is proportional to the linear halo bias. To understand 
better the reasons behind the failure of standard Lagrangian local bias, in which
the halo overdensity is a function of the local mass overdensity only, we explore 
the effect of a primordial bispectrum on the 2-point correlation of discrete 
density peaks. 
We show that the effective local bias expansion to peak clustering vastly 
simplifies the calculation. We generalize this approach to excursion set peaks and 
demonstrate that the resulting non-Gaussian amplitude, which is a weighted sum of 
quadratic bias factors, precisely agrees with the peak-background split expectation, 
which is a logarithmic derivative of the halo mass function with respect to the 
normalisation amplitude. 
We point out that statistics of thresholded regions can be computed using the same 
formalism. Our results suggest that halo clustering statistics can be modelled 
consistently (in the sense that the Gaussian and non-Gaussian bias factors agree 
with peak-background split expectations) from a Lagrangian bias relation only if 
the latter is specified as a set of constraints imposed on the linear density field. 
This is clearly not the case of standard Lagrangian local bias. Therefore, one is 
led to consider additional variables beyond the local mass overdensity.
}

\begin{document}

\maketitle
\flushbottom

\setcounter{footnote}{0}

\section{Introduction}
\label{sec:intro}

Recently, galaxy clustering has become a serious contender to the cosmic microwave 
background (CMB) in the hunt for primordial non-Gaussianity 
(see \cite{desjacques/seljak:2010,verde:2010} for recent reviews). 
Non-Gaussian initial conditions can indeed leave strong scale-dependent signatures 
in the galaxy power spectrum \cite{dalal/etal:2008} and bispectrum 
\cite{scoccimarro/etal:2004}. 
However, one of the main caveats in the theoretical modelling of galaxy clustering 
statistics and the interpretation of large scale surveys data is galaxy biasing: 
unlike CMB temperature measurements, galaxies furnish a distorted picture of the 
primordial curvature perturbations because they preferentially trace overdense regions 
of the Universe. This induces a bias between the galaxy and matter distributions, as 
was first pointed out in \cite{kaiser:1984}.

One of the simplest biasing schemes is the local bias model \cite{fry/gaztanaga:1993}.
In its Lagrangian formulation, galaxies are a Poisson sampling of some continuous 
overdensity field $\delta_g(\mathbi{x})$ which is a function of the local, linear  
matter overdensity $\delta(\mathbi{x})$ only. In general, $\delta_g(\mathbi{x})$
is written as 
$\delta_g(\mathbi{x})=b_1\delta(\mathbi{x})+b_2\delta^2(\mathbi{x})/2+\cdots$, where 
the $b_i$ are the Lagrangian bias parameters.
For Gaussian initial conditions, the predictions of this model are in reasonable 
(though not perfect) agreement with halo clustering statistics extracted from N-body 
simulations once gravitational evolution is taken into account. 
However, for a non-zero bispectrum of primordial curvature perturbations, the local 
bias expansion gives rise to a correction to the linear halo bias whose amplitude is 
proportional to $b_2\fnl$ 
\cite{taruya/koyama/matsubara:2008,sefusatti:2009,jeong/komatsu:2009}, 
whereas N-body simulations clearly indicate that the ampltitude scales according to 
$b_1\fnl$ 
\cite{dalal/etal:2008,desjacques/seljak/iliev:2009,pillepich/porciani/hahn:2010}. 
By constrast, the peak-background split ansatz \cite{kaiser:1984} predicts the 
correct amplitude $\propto b_1$ for the non-Gaussian bias 
\cite{slosar/etal:2008,schmidt/kamionkowski:2010}, as well as an additional, initially 
overlooked correction 
\cite{desjacques/etal:2011a,scoccimarro/etal:2012}
that significantly improves the agreement with N-body simulations 
\cite{desjacques/etal:2011b}. Does this imply that we should give up on local bias 
when it comes to primordial non-Gaussianity? Hopefully not, especially since a working
local bias approach would be extremely useful for the computation of the galaxy 
bispectrum with generic initial conditions \cite{jeong/komatsu:2009,sefusatti:2009}.

Several alternatives to local Lagrangian bias have been explored in the context of 
primordial non-Gaussianity. In the approach of \cite{giannantonio/porciani:2010}, 
which is an extension of \cite{mcdonald:2008}, the halo overdensity field is a 
function of both the local matter density and (non-local) Gaussian part of the 
primordial Bardeen potential $\Phi(\mathbi{x})$. 
However, the non-Gaussian bias acquires additional corrections that strongly depend 
on the smoothing scale $R_l$ (i.e. the cell size, which is distinct from the 
Lagrangian radius $R_s$ of a halo). To cure this problem, \cite{schmidt/etal:2012}
have recently proposed a renormalization procedure that absorbs $R_l$-dependent terms, 
so that the non-Gaussian bias amplitude is effectively independent of $R_l$. 
Furthermore, statistics of thresholded regions can provide insights into the 
relationship between local bias and peak-background split \cite{matsubara:2012}. 
In the specific case of sharp clipping, both peak-background split 
and the corresponding local series expansion consistently predict the non-Gaussian 
halo bias \cite{ferraro/etal:2012}. 
However, one of the main drawbacks of this model is its poor fit to the clustering 
of dark matter halos with Gaussian initial conditions. Clearly, it is essential 
that the model be able to reproduce Gaussian clustering statistics as well.

In this work, we compute the non-Gaussian correction to the linear bias within the 
peak formalism \cite{bardeen/etal:1986}. The central assumption of this model is 
the correspondence between virialized halos and maxima of the initial density field,
so that the biased tracers constitute a genuine point set. 
Another essential difference with most of the approaches mentioned above (thresholded 
regions excepted) lies in the fact that clustering statistics of initial density peaks 
are fully specificed upon enforcing the peak constraint, without the need for local 
bias expansion or peak-background split. 
Yet another motivation for our work is the non-universality of the multiplicity function 
of initial density peaks. Non-universality of the mass function must be accounted for 
to obtain exquisite agreement with measurements from N-body simulations 
\cite{scoccimarro/etal:2012}.

The outline of the paper is as follows. 
First, we demonstrate that the non-Gaussian bias obtained from a direct calculation 
of the non-Gaussian 2-point probability density agrees with that derived from the 
effective, local bias expansion advocated by \cite{desjacques:2012}. This shows that 
the latter can be extended to non-Gaussian initial conditions. 
Second, we prove that the resulting non-Gaussian bias agrees with that inferred from 
peak-background split, in which primordial non-Gaussianity rescales the amplitude of 
density fluctuations $\sigma_8$ in a scale-dependent manner. 
Third, we generalize these results to excursion set peaks \cite{paranjape/sheth:2012} 
and, in particular, provide a local bias expansion along the lines of 
\cite{desjacques:2012} that includes the first-crossing condition.
We find again consistency with peak-background split expectations. This is especially 
interesting since excursion set peaks have recently been shown to give a very good 
match to the Gaussian mass function and linear bias of dark matter halos 
\cite{paranjape/sheth/desjacques:2012}. 
Finally, we discuss our results in light of recent work before summarizing our 
findings.

\section{A brief overview of the peak formalism}

We are interested in the 2-point correlation function of density maxima of the 
initial density field $\delta(\mathbi{x})$, which may not follow perfect Gaussian
statistics. The assumption is that nonlinear structures form from the gravitational 
collapse of local maxima of the linear, smoothed overdensity field  $\delta_s(\mathbi{x})$. 
To define a local maximum, one must require that the first derivative vanish and 
the second derivatives be negative. Following \cite{bardeen/etal:1986}, we normalize 
$\delta_s(\mathbi{x})$ and its derivatives as follows:
\begin{equation}
\nu(\mathbi{x}) \equiv \frac{1}{\sigma_0}\delta_s(\mathbi{x}) \, , \quad
\eta_i(\mathbi{x}) \equiv \frac{1}{\sigma_1}\partial_i\delta_s(\mathbi{x}) \, , \quad
\zeta_{ij}(\mathbi{x}) \equiv \frac{1}{\sigma_2}\partial_i\partial_j\delta_s(\mathbi{x}) \;.
\end{equation}
Here, the spectral moments $\sigma_j$ are given by
\begin{equation}
\label{eq:spectral}
\sigma_j^2 \equiv \int\!\! \frac{d^3k}{(2\pi)^3}\, P_s(k) k^{2j} = \frac{1}{2\pi^2}
\int_0^\infty\!\!dk\,k^{2(j+1)} P_s(k)\, ,
\end{equation}
where $P_s(k)=W^2(k R_s)P_\delta(k)$ is the power spectrum of the initial density field 
smoothed on the Lagrangian scale $R_s$ of a halo with a spherically symmetric kernel
$W(k R_s)$. 
Density maxima are a point process and, therefore, their abundance is formally a sum 
of Dirac distributions. 
On using the Kac-Rice formula \cite{kac:1943,rice:1945} and enforcing the peak 
constraint, the number density of peaks of height $\nu_c=\delta_c/\sigma_0$, where
$\delta_c$ is the critical threshold for spherical collapse \cite{gunn/gott:1972}, 
in a 3-dimensional Gaussian random field can be written as \cite{bardeen/etal:1986}
\begin{equation}
\label{eq:npeak}
\npk(\mathbi{x};\nu_c) = \frac{3^{3/2}}{R_\star^3} \left|\det\pmb{\zeta}\right| 
\delta_D\!\left(\pmb{\eta}\right) \theta_H\!\left(\lambda_3\right) 
\delta_D\!\left(\nu-\nu_c\right) \;.
\end{equation}
where $R_\star\equiv\sqrt{3}\sigma_1/\sigma_2$ is the characteristic radius of peaks and 
$\lambda_3$ is the smallest eigenvalue of the matrix $-\zeta_{ij}$, so that the term 
$\theta_H(\lambda_3)$ ensures that we are only counting density maxima. Here and 
henceforth, we will omit the dependence of $\npk(\vx;\nu_c)$ on $R_s$ for concisencess,
but one should bear in mind that $\npk$ depends explicitly on both $\nu_c$ and $R_s$. 
Note also that dealing with peaks of the density field does not imply $\nabla\Phi=0$ 
at the peak positions. This assumption was sometimes invoked in the derivation of the
non-Gaussian bias \cite{dalal/etal:2008,giannantonio/porciani:2010}

The $n$-point correlation function of peaks of given significance $\nu_c$ is defined 
as the ensemble average
\begin{equation}
\label{def:npointcorr}
1+\xi_\text{pk}^{(n)}(\mathbi{x}_1,\mathbi{x}_2,\cdots\mathbi{x}_n;\nu_c) \equiv 
\frac{\left\langle \npk(\mathbi{x}_1;\nu_c)\npk(\mathbi{x}_2;\nu_c) \cdots 
\npk(\mathbi{x}_n;\nu_c) \right\rangle}{\left\langle\npk(\mathbi{x};\nu_c)\right\rangle^n} \, , 
\end{equation}
where the average is taken with respect to the derivatives of $\delta_s(\mathbi{x})$,
\begin{align}
\label{eq:fullNpoint}
\left\langle \prod_i \npk(\mathbi{x}_i;\nu_c) \right\rangle = & 
\left( \frac{3^{3/2}}{R_\star^3} \right)^n \left\langle \prod_i 
\left|\det\pmb{\zeta}_i\right| \delta_D\!\left(\pmb{\eta}_i\right) 
\theta_H\!\left[\lambda_3(i)\right]\delta_D\!(\nu_i-\nu_c) \right\rangle \\
= & \left( \frac{3^{3/2}}{R_\star^3} \right)^n \prod_i \int d^3\eta_i d^6\zeta_i 
\left|\det\pmb{\zeta}_i\right| \delta_D\!\left(\pmb{\eta}_i\right) 
\theta_H\!\left[\lambda_3(i)\right] P_n(\mathbi{x}_i;\nu_i\equiv\nu_c,\pmb{\eta}_i,\pmb{\zeta}_i) 
\nonumber \;.
\end{align}
Here, $P_n(\mathbi{x}_i;\nu_i,\pmb{\eta}_i,\pmb{\zeta}_i)$ is the joint probability distribution 
for the variables $(\nu,\pmb{\eta},\pmb{\zeta})$ at $n$ different spatial locations. 
The $n$-point connected or irreducible correction function is defined in such a way that
$\left\langle \npk(\mathbi{x}_1;\nu_c)\cdots\npk(\mathbi{x}_n;\nu_c) \right\rangle$ 
is a sum of terms, each of which pertains to a partition of the set
of $n$ locations $\mathbi{x}_1,\cdots\mathbi{x}_n$ (e.g. \cite{peebles:1980,bernardeau:1992}).
In this paper, we will focus on the connected 2-point correlation of peaks of height $\nu_c$ 
(at a separation $r=|\mathbi{x}_2-\mathbi{x}_1|$)
\begin{equation}
\xi_\text{pk}^{(2)}(r;\nu_c)\equiv 
\frac{\left\langle \npk(\mathbi{x}_1;\nu_c)\npk(\mathbi{x}_2;\nu_c) \right\rangle}
{\left\langle\npk(\mathbi{x};\nu_c)\right\rangle^2}-1 \;. 
\end{equation}
This 2-point correlation function can be evaluated in two different ways, either from an 
explicit computation of $\left\la\npk(\mathbi{x}_1;\nu_c)\npk(\mathbi{x}_2;\nu_c)\right\ra$ 
for Gaussian initial conditions, see \cite{bardeen/etal:1986,regos/szalay:1995,
desjacques:2008,desjacques/etal:2010}) or from the much simpler local bias approach 
proposed by \cite{desjacques:2012}. Reassuringly, the two methods agree, yet the second 
is much faster and physically more intuitive.

\section{Peak 2-point correlation with non-Gaussian initial conditions}
\label{sec:pkngbias}

In this Section, we derive the non-Gaussian corrections to the 2-point correlation 
function $\xpk(r;\nu_c)$ in two different ways: from a direct computation of the 
joint probability density $P_2(\vx_i;\nu_i,\pmb{\eta}_i,\pmb{\zeta}_i)$ and from the 
effective local bias approach \cite{desjacques:2012}. We describe the two methods 
and demonstrate that they lead to the same non-Gaussian bias correction.

\subsection{From an Edgeworth expansion of the probability distributions}

In the presence of non-Gaussian initial conditions the generating function formalism 
can be applied to find an explicit expression for the joint probability 
$P_n(\mathbi{x}_i;\nu_i,\pmb{\eta}_i,\pmb{\zeta}_i)$ before computing correlators of 
$\npk(\mathbi{x},\nu_c)$. This functional approach has been used in, e.g. 
\cite{politzer/wise:1984,grinstein/wise:1986,matarrese/etal:1986,matsubara:2003}. 
Here, we briefly review the basic arguments in the form appropriate for our purpose.

\subsubsection{Edgeworth formalism}

Let $P(y_1,y_2,\cdots,y_N)$ be a generic $N$-dimensional probability distribution 
function. The {\em connected} correlation functions of the random variables $y_i$ 
can be written in terms of the generating function $Z(J_1,J_2,\cdots J_N) \equiv Z(J)$,
which is defined as
\begin{equation}
\label{eq:Z}
Z(J) = \int dy_1 dy_2\cdots dy_N P(y_1,y_2,\cdots y_N) 
e^{iJ_1y_1 + iJ_2y_2 + \cdots + iJ_Ny_N} 
\equiv \left\langle e^{iJ_iy_i} \right\rangle \;.
\end{equation}
The connected correlators are then given by derivatives of the logarithm of $Z(J)$,
\begin{equation}
\xi_{m_1 m_2\cdots m_N}^{(c)} \equiv 
\left\langle y_1^{m_1} y_2^{m_2}\cdots y_N^{m_N} \right\rangle_{(c)} 
= \left. \frac{1}{i^n} \frac{d^{m_1+\cdots+m_N}}{dJ_1^{m_1} dJ_2^{m_2}\cdots dJ_N^{m_N}} 
\log Z(J) \right|_{J_1=J_2=\cdots= J_N = 0} \; .
\end{equation}
Note that the $y_i$ may be defined at different spatial locations, but this is not 
essential to our discussion.
The joint probability $P(y_1,\cdots,y_N)$ can then be constructed from the generating 
function $Z(J)$. For random variables with zero expectations values, we have
\begin{align}\label{P4}
P(y_1,y_2,\cdots,y_N) &= \exp \left[ \sum_{n=3}^\infty (-1)^n \!\!
\sum_{m_1,\cdots m_N} \frac{\xi_{m_1\cdots m_N}^{(c)}}{m_1!\cdots m_N!} 
\left( \frac{\partial}{\partial y_1} \right)^{m_1}
\cdots\left(\frac{\partial}{\partial y_N}\right)^{m_N} \right]
\nonumber\\
& \qquad \times \frac{1}{\sqrt{(2\pi)^N\det\vmm}} 
\exp \left[ -\frac{1}{2}\sum_{i,j} y_i\left(\vmm^{-1}\right)_{ij}y_j \right] \, ,
\end{align}
where $\sum_i m_i=N$ and $\vmm_{ij} \equiv \langle y_i y_j \rangle$ is the covariance 
matrix. We can replace the derivatives with respect to $y_i$ by introducing multi-variate 
Hermite polynomials defined as
\begin{equation}
H_{\mathbi{m}}(\mathbi{y};\vmm^{-1}) \equiv 
(-1)^n \exp \left( \frac{1}{2}\mathbi{y}^T\vmm^{-1}\mathbi{y} \right) 
\left( \frac{\partial}{\partial y_1} \right)^{m_1} \cdots 
\left( \frac{\partial}{\partial y_N} \right)^{m_N} 
\exp \left( -\frac{1}{2}\mathbi{y}^T\vmm^{-1}\mathbi{y} \right) \, ,
\end{equation}
where $\mathbi{y}=(y_1,\cdots,y_N)$ and $\mathbi{m} = (m_1,m_2,\cdots m_N)$. On expanding 
the exponential factor in the correlations $\xi_{m_1\cdots m_N}^{(c)}$ with $n\geq 3$ 
(assuming they are small relative to the second order moments), we can eventually write down 
the joint probability density as
\begin{align}
P(y_1,y_2\cdots,y_N) = & \frac{1}{\sqrt{(2\pi)^N\det\vmm}} 
\exp \left[ -\frac{1}{2}\sum_{i,j} y_i\left(\vmm^{-1}\right)_{ij}y_j \right]
\nonumber\\
& \times \left[ 1 + \sum_{n=3}^\infty\sum_{m_1,\cdots m_N} 
\frac{\xi_{m_1\cdots m_N}^{(c)}}{m_1!\cdots m_N!} H_{\mathbi{m}}(\mathbi{y};\vmm^{-1}) 
+ \cdots \right]  \, .
\label{eq:npointnGprob}
\end{align}
The first line on the right-hand side is a multivariate Gaussian, whereas the second line 
is the non-Gaussian correction represented as an Edgeworth series in the $n\geq 3$ connected 
correlation functions. This series expansion can then be substituted in place of $P_n$ in 
(\ref{eq:fullNpoint}) to evaluate the non-Gaussian contributions to the $N$-point 
function.

\subsubsection{Non-Gaussian bias }

We restrict ourselves to the corrections from the non-Gaussian sector of our interest, 
i.e. the contributions from 3-point correlation functions including the derivatives of 
the density field. 
A total of eight non-vanishing, distinct combinations are possible, not including 
the exchange of the coordinates $\mathbi{x}_1 \leftrightarrow \mathbi{x}_2$. These are
\begin{equation}
\begin{split}
\left\la\nu^2(\mathbi{x}_1)\nu(\mathbi{x}_2)\right\ra, \quad &
\left\la\nu^2(\mathbi{x}_1)\zeta_{ij}(\mathbi{x}_2)\right\ra, 
\\
\left\la\nu(\mathbi{x}_1)\zeta_{ij}(\mathbi{x}_1)\nu(\mathbi{x}_2)\right\ra, \quad &
\left\la\nu(\mathbi{x}_1)\zeta_{ij}(\mathbi{x}_1)\zeta_{lm}(\mathbi{x}_2)\right\ra, 
\\
\left\la\eta_i(\mathbi{x}_1)\eta_j(\mathbi{x}_1)\nu(\mathbi{x}_2)\right\ra, \quad &
\left\la\eta_i(\mathbi{x}_1)\eta_j(\mathbi{x}_1)\zeta_{lm}(\mathbi{x}_2)\right\ra \, , 
\\
\left\la\zeta_{ij}(\mathbi{x}_1)\zeta_{lm}(\mathbi{x}_1)\nu(\mathbi{x}_2)\right\ra, \quad &
\left\la\zeta_{ij}(\mathbi{x}_1)\zeta_{kl}(\mathbi{x}_1)\zeta_{mn}(\mathbi{x}_2)\right\ra \;.
\end{split}
\label{eq:3ptcorrel}
\end{equation}
These 3-point correlators can generically be written as 
$\left\langle T_1\delta_s(\mathbi{x}_1) T_2\delta_s(\mathbi{x}_1) T_3\delta_s(\mathbi{x}_2)
\right\rangle$, where $T_i$ are differential operators. For instance, setting $T_i=-\nabla^2$ 
yields the curvature 
\begin{equation}
u(\mathbi{x})\equiv -\frac{1}{\sigma_2}\nabla^2\delta_s(\mathbi{x})
\end{equation} 
of the smoothed density field. After some algebra, the Fourier transform of this 3-point 
correlator is
\begin{align}
& \int\!\!d^3\mathbi{r}\, \left\langle T_1\delta_s(\mathbi{x}_1) T_2\delta_s(\mathbi{x}_1) 
T_3\delta_s(\mathbi{x}_2)\right\rangle e^{-i\mathbi{k}\cdot\mathbi{r}} 
\nonumber\\
& \qquad =
\int\!\!\frac{d^3\mathbi{q}}{(2\pi)^3}\, T_1(\mathbi{q}) T_2(-\mathbi{k}-\mathbi{q}) T_3(\mathbi{k}) 
\calM_s(q)
\calM_s(|\mathbi{k}+\mathbi{q}|) \calM_s(k) \xi_\Phi^{(3)}\!(\mathbi{q},-\mathbi{k}-\mathbi{q},
\mathbi{k}) \, .
\end{align}
Here, the 3-point function $\xi_\Phi^{(3)}(\mathbi{k}_1,\mathbi{k}_2,\mathbi{k}_3)$ is that of 
$\Phi(\mathbi{x})$, and $\calM_s(k)=\calM(k) W(kR_s)$ where ${\cal M}(k)$ is the transfer 
function between $\Phi(\mathbi{x})$ and the linear, smoothed density field at redshift $z$.
Here and henceforth, we will only consider the limit $k\to 0$, in which the right-hand side is 
approximately
\begin{equation}
\label{eq:squeezed1}
T_3(k) \calM_s(k) P_\phi(k) \int\!\!\frac{d^3\mathbi{q}}{(2\pi)^3}\, T_1(\mathbi{q}) T_2(-\mathbi{q})
\calM_s^2(q) \frac{\xi_\Phi^{(3)}\!(\mathbi{q},-\mathbi{q},-\mathbi{k})}{P_\phi(k)}\;,
\end{equation}
but note that our analysis remains valid outside the squeezed limit of the 3-point correlators.
Clearly, this expression scales proportionally to $k^{-2}$ only if $T_3(k)\equiv 1$ or,
equivalently, if $T_3\delta_s(\mathbi{x}_2)\equiv \nu(\mathbi{x}_2)$.
In the specific case of the constant-$\fnl\phi^2$ model, the operators $T_i(\pm\mathbi{q})$ 
give rise to the spectral index $\sigma_1^2$ or $\sigma_2^2$, depending on their detailed form.
In the general case, the scale-dependence and amplitude of the non-Gaussian bias correction
is governed by the squeezed limit of the bispectrum of curvature perturbations. Accounting for
a possible scale-dependence of $\fnl$, we parametrize the latter as 
\cite{desjacques/etal:2011a,schmidt/etal:2012}
\begin{equation}
\label{eq:squeezed2}
\xi_\Phi^{(3)}\!(\mathbi{q},\mathbi{k}-\mathbi{q},-\mathbi{k}) \stackrel{k\rightarrow0}{=}
2 A \fnl(k_p) \biggl(\frac{q}{k_p}\biggr)^{2\alpha_1}\biggl(\frac{q}{k}\biggr)^{2\alpha_2} 
P_\phi(q) P_\phi(k) \;.
\end{equation}
Here, $k_p$ is some reference wavenumber. This parametrization encompasses many bispectrum
shapes considered in the literature.
For instance, the quadratic coupling $\fnl\phi^2$ with $k$-dependent $\fnl(k)\propto k^{n_f}$ 
considered in \cite{shandera/etal:2011} (see also \cite{byrnes/gong:2013}) 
corresponds to $A=1$, $\alpha_1=n_f/2$ and $\alpha_2=0$,
 whereas the equaliteral bispectrum shape \cite{creminelli/etal:2006} yields $A=3/2$, 
$\alpha_1=0$ and $\alpha_2=(n_s-4)/3$. Consequently, (\ref{eq:squeezed1}) simplifies to
\begin{equation}
T_3(k) \calM_s(k) P_\phi(k) 
2 A \frac{\fnl(k_p)}{k_p^{2\alpha_1}} k^{-2\alpha_2} 
\int\!\!\frac{d^3\mathbi{q}}{(2\pi)^3}\, q^{2(\alpha_1+\alpha_2)} T_1(\mathbi{q}) 
T_2(-\mathbi{q}) \calM_s^2(q) P_\phi(q) \, ,
\end{equation}
and is proportional to the square $\sigma_j^2$ of a spectral moment (\ref{eq:spectral}) 
with $j$ determined by the values of $\alpha_1$, $\alpha_2$ and the scaling of the Fourier 
transforms $T_1(\mathbi{q})$ and $T_2(\mathbi{q})$.

The calculation of the non-Gaussian corrections to $\xpk(r;\nu_c)$ from the Edgeworth expansion
is straightforward.
Including symmetric numerical factors and denoting $\alpha\equiv \alpha_1+\alpha_2$, the 
contributions of the 3-point correlators given in (\ref{eq:3ptcorrel}) to the 2-point power 
spectrum of discrete density peaks are, formatted as in (\ref{eq:3ptcorrel}),
\begin{equation}
\begin{split}
b_{20}(1) b_{10}(2)\sigma_\alpha^2\, , \quad &
b_{20}(1) b_{01}(1) k^2 \sigma_\alpha^2\, , 
\\
2 b_{11}(1) b_{10}(2) \sigma_{\alpha+1}^2\, , \quad &
2 b_{11}(1) b_{01}(2) k^2 \sigma_{\alpha+1}^2\, , 
\\
-3 b_{10}(2)\sigma_{\alpha+1}^2\, , \quad &
-3 b_{01}(2) k^2 \sigma_{\alpha+1}^2 \, ,
\end{split}
\label{eq:3ptpk1}
\end{equation}
and
\begin{gather}
\biggl[b_{02}(1)-5-2\partial_\alpha\log G_0^{(\alpha)}\!(\gamma_1,\gamma_1\nu)
\Bigr\lvert_{\alpha=1}\biggr] b_{10}(2) \sigma_{\alpha+2}^2 \, ,
\nonumber \\
\biggl[b_{02}(1)-5-2\partial_\alpha\log G_0^{(\alpha)}\!(\gamma_1,\gamma_1\nu)
\Bigr\lvert_{\alpha=1}\biggr] b_{01}(2) k^2 \sigma_{\alpha+2}^2 \;,
\label{eq:3ptpk2}
\end{gather}
where we have omitted the common multiplicative factor 
$4 A \fnl(k_p) k_p^{-2\alpha_1} k^{-2\alpha_2} \calM_s(k) P_\phi(k)$ for brevity.
Explicit expressions for the linear and quadratic bias factors $b_{ij}$ are given by
\cite{desjacques:2008,desjacques/etal:2010,desjacques:2012}
\begin{gather}
b_{10} = \ov{b}_\nu \,,\qquad b_{01} = \ov{b}_u \,, 
\nonumber\\
b_{20} = \ov{b_\nu^2}-\frac{1}{\sigma_0^2(1-\gamma_1^2)} \,, \qquad
b_{11} = \ov{b_\nu b_u}+\frac{\gamma_1^2}{\sigma_1^2(1-\gamma_1^2)} \,,\qquad
b_{02} = \ov{b_u^2}-\frac{1}{\sigma_2^2(1-\gamma_1^2)} \;.
\label{eq:bias-bij}
\end{gather}
Here, $\gamma_1\equiv \sigma_1^2/(\sigma_0\sigma_2)$ is a dimensionless number that 
takes values between zero and unity depending on the shape of the smoothed density 
power spectrum \cite{bardeen/etal:1986}. Moreover,
\begin{equation}
\label{eq:bvbu}
b_\nu = \frac{1}{\sigma_0}\left(\frac{\nu_c-\gamma_1 u}{1-\gamma_1^2}\right) \,, \qquad
b_u = \frac{1}{\sigma_2}\left(\frac{u-\gamma_1\nu_c}{1-\gamma_1^2}\right) \, ,
\end{equation}
where the overline designates the average over $u$ at the locations of density peaks. The 
function $G_n^{(\alpha)}\!(\gamma_1,\gamma_1\nu_c)$ is defined in \cite{desjacques/etal:2010},
\begin{align}
G_n^{(\alpha)}(\gamma,z) & = 
\int_0^\infty du \, u^n \frac{f(u,\alpha)}{\sqrt{2\pi\left(1-\gamma^2\right)}}  
\exp\left[ -\frac{(u-z)^2}{2\left(1-\gamma\right)^2} \right] \nonumber \, ,
\\
f(u,\alpha) & = \frac{3^25^{5/2}}{\sqrt{2\pi}} 
\left( \int_0^{u/4} dv \int_{-v}^v dw + \int_{u/4}^{u/2} dv \int_{3v-u}^v dw \right) 
F(u,v,w) e^{-5\alpha \left( 3v^2+w^2 \right)/5} \nonumber \, ,
\\
\label{eq:shapefactor}
F(u,v,w) & = (u-2w) \bigl[ (u+w)^2-9v^2 \bigr] v \left( v^2-w^2 \right) \, .
\end{align}
Furthermore, although the bias parameters $b_{ij}$ do not depend on position, we have added
an argument of (1) or (2) to indicate whether they were evaluated from variables at position 
$\mathbi{x}_1$ or $\mathbi{x}_2$. Adding up all the terms, we eventually obtain
\begin{align}
\label{DeltaP_Edgeworth}
\Delta P_{\rm pk}(k) &= 4A \, \frac{\fnl(k_p)}{k_p^{2\alpha_1}}\, k^{-2\alpha_2} 
\calM_s(k) P_\phi(k) \left(b_{10}+b_{01}k^2\right) \\
& \quad \times
\biggl\{ \sigma_\alpha^2 b_{20} + 2 \sigma_{\alpha+1}^2 b_{11} + \sigma_{\alpha+2}^2 b_{02}
-3\sigma_{\alpha+1}^2 - \Bigl[ 5+2\,\partial_\alpha\log G_0^{(\alpha)}\!(\gamma_1,\gamma_1\nu)
\Bigr\lvert_{\alpha=1} \Bigr] \sigma_{\alpha+2}^2 \biggr\} \nonumber \;.
\end{align}
We recognize the linear peak bias $c_1(k)\equiv (b_{10}+b_{01}k^2) W(k R_s)$ and, inside the 
square brackets, the quadratic bias factors $\chi_{10}$ and $\chi_{01}$ defined by 
\cite{desjacques:2012},
\begin{equation}
\label{eq:bias-chi}
\chi_{10}= -\frac{3}{2\sigma_1^2} \,, \qquad
\chi_{01}= -\frac{5}{2\sigma_2^2}
\biggl[1+\frac{2}{5}\partial_\alpha\log G_0^{(\alpha)}\!(\gamma_1,\gamma_1\nu_c)
\Bigr\lvert_{\alpha=1}\biggr] \;.
\end{equation}
We can read off the (scale-dependent) non-Gaussian correction $\Delta c_1(k)$ to the linear 
peak bias from $\Delta P_{\rm pk}(k)=2 c_1(k)\Delta c_1(k) P_\delta(k)$.
Notice that, in order to highlight the connection with the formalism of Matsubara 
\cite{matsubara:2012}, we have adopted his $c_N$-notation for the renormalized bias functions. 
However, we emphasize that our calculation does not involve any renormalization scheme 
whatsoever.

\subsection{From a local bias approach to peak clustering}

\subsubsection{Effective local bias expansion}

As shown in \cite{desjacques:2012}, the 2-point correlation function $\xpk(r;\nu_c)$ of 
peaks of a Gaussian random field can be thought of, up to second order at least, as arising 
from the {\it effective} local bias expansion
\begin{align}
\dpk(\mathbi{x}) &= \sigma_0 b_{10} \nu(\mathbi{x}) +\sigma_2 b_{01} u(\mathbi{x}) 
+ \frac{1}{2} \sigma_0^2 b_{20}\nu^2(\mathbi{x}) 
+ \sigma_0 \sigma_2 b_{11}\nu(\mathbi{x}) u(\mathbi{x}) 
+ \frac{1}{2} \sigma_2^2 b_{02} u^2(\mathbi{x}) 
\nonumber \\
& \quad 
+ \sigma_1^2 \chi_{10} \eta^2(\mathbi{x}) + \sigma_2^2 \chi_{01} \zeta^2(\mathbi{x}) \, ,
\label{eq:dpk}
\end{align}
provided that we ignore all the contributions involving powers of zero lag moments.
Here, $\eta^2(\mathbi{x})$ and $\zeta^2(\mathbi{x})$ are the square length of the vector of 
first derivatives and the trace of the squared traceless part of the hessian matrix 
$\partial_i\partial_j\delta_s(\mathbi{x})$ respectively, 
\begin{equation}
\label{eq:expparameters}
\eta^2(\mathbi{x}) \equiv \frac{1}{\sigma_1^2}\bigl[\grad\delta_s(\mathbi{x})\bigr]^2 \,, \qquad
\zeta^2(\mathbi{x}) \equiv \frac{1}{\sigma_2^2}
\biggl[ \Bigl( \partial_i\partial_j-\frac{1}{3}\delta_{ij}\Delta \Bigr) \delta_s(\mathbi{x})
\biggr]^2 \, .
\end{equation}
This expansion is local except for the filtering of the mass density field. It is 
effective in the sense that $\dpk(\mathbi{x})$ is not a count-in-cell overabundance of peaks, 
but merely some idealized continuous overdensity field that can be used to derive 
$\xpk(r;\nu_c)$ without having to compute the joint probability density $P_2$ that describes
covariances between variables defined at two different locations.

The peak bias factors $b_{ij}$ and $\chi_{ij}$ can be derived from a peak-background split 
argument where the long-wavelength perturbation shifts the mean of the probability distributions 
characterizing the rotational invariants $\nu(\mathbi{x})$, $u(\mathbi{x})$, $\eta^2(\mathbi{x})$ 
and $\zeta^2(\mathbi{x})$.
The variables $\nu$ and $u$ are distributed normally, whereas $3\eta^2$ and $5\zeta^2$ are 
distributed as chi-squared ($\chi^2$) variates with 3 and 5 degrees of freedom, respectively.
The resulting non-central probability densities can be expanded in sets of orthogonal
polynomials which, on enforcing the peak constraints, yield the peak bias parameters. 
The orthogonal polynomials associated with the correlated variables $\nu$ and $u$ are 
bivariate Hermite polynomials $H_{ij}$, whereas those pertaining to $\eta^2(\mathbi{x})$ and 
$\zeta^2(\mathbi{x})$ are generalized Laguerre polynomials $L_k^{(\alpha)}$. We thus have
\begin{equation}
\begin{split}
b_{ij} &= \frac{1}{\sigma_0^i\sigma_2^j\bnpk}
\int\!\!d\nu d^3\pmb\eta d^6\pmb\zeta \, \npk(\mathbi{x};\nu_c) 
H_{ij}(\nu,u)\, P_1(\mathbi{x};\nu,\pmb\eta,\pmb\zeta) \, , 
\\
\chi_{k0} &= \frac{(-1)^k}{\sigma_1^{2k}\bnpk}
\int\!\!d\nu d^3\pmb\eta d^6\pmb\zeta\, \npk(\mathbi{x};\nu_c)\,
L_k^{(1/2)}\!\!\left(\frac{3\eta^2}{2}\right) P_1(\mathbi{x};\nu,\pmb\eta,\pmb\zeta) \, ,
\\
\chi_{0k} &= \frac{(-1)^k}{\sigma_2^{2k}\bnpk}
\int\!\!d\nu d^3\pmb\eta d^6\pmb\zeta\, \npk(\mathbi{x};\nu_c)\,
L_k^{(3/2)}\!\!\left(\frac{5\zeta^2}{2}\right) P_1(\mathbi{x};\nu,\pmb\eta,\pmb\zeta) \;.
\end{split}
\end{equation}
The probability density $P_1$ can be factorized into the product of a bivariate Gaussian 
${\cal N}(\nu,u)$ times $\chi_3^2(3\eta^2)$ and $\chi_5^2(5\zeta^2)$, where $\chi_k^2(x)$ 
is a $\chi^2$-distribution with $k$ degrees of freedom [see also (\ref{ESP1pointP})].

The bias factors $b_{ij}$ can be easily computed from the generating function given in 
\cite{desjacques:2012}. In particular, the linear and quadratic bias factors associated to 
the scalars $\nu$ and $u$ are given by (\ref{eq:bias-bij}), whereas those pertaining to the 
quadratic invariants are given by (\ref{eq:bias-chi}). 
Note that the term $\chi_{01}\zeta^2(\mathbi{x})$ in the effective local series expansion 
has a simple physical meaning: it represents the leading-order contribution of asphericity 
in the peak density profiles to the clustering of density peaks. Therefore, it vanishes in 
the high peak limit $\nu\gg 1$ since high peaks are close to spherical.

\subsubsection{Non-Gaussian bias}

From the effective local bias expansion (\ref{eq:dpk}), the leading order contribution to 
the non-Gaussian peak 2-point correlation is immediately computed as
\begin{align}
\Delta\xpk(r;\nu_c) &= 2\sigma_0^3 b_{10} b_{20} \la\nu_1^2 \nu_2\ra +4\sigma_0^2\sigma_2 
b_{10} b_{11}\la\nu_1 u_1\nu_2\ra + 2\sigma_0\sigma_2^2 b_{10} b_{02}\la u_1^2\nu_2\ra
\nonumber \\
& \quad 
+ 4\sigma_0\sigma_1^2\chi_{10} b_{10}\la\eta_1^2\nu_2\ra + 4\sigma_0\sigma_2^2
\chi_{01} b_{10}\la\zeta_1^2\nu_2\ra \nonumber \\
& \quad
+ \big( \sigma_0, b_{10}, \nu_2 \to \sigma_2, b_{01}, u_2 \big) \, .
\end{align}
The last-line in the right-hand side signifies that the other terms can be obtained 
upon replacing all occurences of $(\sigma_0, b_{10}, \nu_2)$ by $(\sigma_2, b_{01}, u_2)$. 
In the low-$k$ limit, the various contributions to the non-Gaussian peak power spectrum 
are given by
\begin{equation}
\begin{split}
\int\!\!d^3\mathbi{r}\,\bigl\la\nu^2(\mathbi{x}_2)\nu(\mathbi{x}_1)\bigr\ra\, 
e^{-i\mathbi{k}\cdot\mathbi{r}} 
~&\stackrel{k\rightarrow0}{=}~ 
\frac{2}{\sigma_0}\left(\frac{\sigma_\alpha}{\sigma_0}\right)^2 
A\, \frac{\fnl(k_p)}{k_p^{2\alpha_1}} k^{-2\alpha_2} \calM_s(k) P_\phi(k) 
\\
\int\!\!d^3\mathbi{r}\,\bigl\la u^2(\mathbi{x}_2)\nu(\mathbi{x}_1)\bigr\ra\, 
e^{-i\mathbi{k}\cdot\mathbi{r}} 
~&\stackrel{k\rightarrow0}{=}~
\frac{2}{\sigma_0}\left(\frac{\sigma_{\alpha+2}}{\sigma_2}\right)^2
A\, \frac{\fnl(k_p)}{k_p^{2\alpha_1}} k^{-2\alpha_2}\calM_s(k) P_\phi(k) 
\\
\int\!\!d^3\mathbi{r}\,\bigl\la u(\mathbi{x}_2)\nu(\mathbi{x}_2)\nu(\mathbi{x}_1)\bigr\ra\, 
e^{-i\mathbi{k}\cdot\mathbi{r}} 
~&\stackrel{k\rightarrow0}{=}~
\frac{2}{\sigma_0}\left(\frac{\sigma_{\alpha+1}}{\sigma_0\sigma_2}\right)^2
A\, \frac{\fnl(k_p)}{k_p^{2\alpha_1}} k^{-2\alpha_2} \calM_s(k) P_\phi(k) 
 \\
\int\!\!d^3\mathbi{r}\,\bigl\la \eta^2(\mathbi{x}_2)\nu(\mathbi{x}_1)\bigr\ra\, 
e^{-i\mathbi{k}\cdot\mathbi{r}} 
~&\stackrel{k\rightarrow0}{=}~
\frac{4}{\sigma_0}\left(\frac{\sigma_{\alpha+1}}{\sigma_1}\right)^2
A\, \frac{\fnl(k_p)}{k_p^{2\alpha_1}} k^{-2\alpha_2} \calM_s(k) P_\phi(k) 
 \\
\int\!\!d^3\mathbi{r}\,\bigl\la \zeta^2(\mathbi{x}_2)\nu(\mathbi{x}_1)\bigr\ra\, 
e^{-i\mathbi{k}\cdot\mathbi{r}} 
~&\stackrel{k\rightarrow0}{=}~
\frac{2}{\sigma_0}\left(\frac{\sigma_{\alpha+2}}{\sigma_2}\right)^2
A\, \frac{\fnl(k_p)}{k_p^{2\alpha_1}} k^{-2\alpha_2} \calM_s(k) P_\phi(k) \;.
\end{split}
\end{equation}
The 5 correlators that involve $u(\mathbi{x}_1)$ instead of $\nu(\mathbi{x}_1)$ yield an extra 
multiplicative factor of $k^2$. On summing all contributions, the leading-order 
non-Gaussian correction to the peak power spectrum reads
\begin{align}
\Delta P_{\rm pk}(k) &= 4A\, \frac{\fnl(k_p)}{k_p^{2\alpha_1}}\, k^{-2\alpha_2} 
\calM_s(k) P_\phi(k) \left(b_{10}+b_{01}k^2\right) \nonumber \\
& \qquad \times
\Bigl[\sigma_\alpha^2 b_{20}+2\sigma_{\alpha+1}^2 b_{11}+\sigma_{\alpha+2}^2 b_{02}
+2\sigma_{\alpha+1}^2\chi_{10}+2\sigma_{\alpha+2}^2\chi_{01}\Bigr] \, .
\end{align}
This result agrees exactly with (\ref{DeltaP_Edgeworth}) obtained from the Edgeworth expansion 
of the joint probability distribution $P_2(\mathbi{x}_i;\nu_i,\pmb{\eta}_i,\pmb{\zeta}_i)$. 
We shall now establish a connection with the peak-background split ansatz.

\section{A peak-background split interpretation of the results}
\label{sec:pbs}

\subsection{Peak-background split and non-Gaussian bias}

As shown in, e.g. \cite{slosar/etal:2008,schmidt/kamionkowski:2010,desjacques/etal:2011a,
smith/etal:2012,scoccimarro/etal:2012}, non-Gaussian corrections to the bias parameters 
can be computed using a peak-background split
\cite{kaiser:1984,bardeen/etal:1986}. 
In its simplest expression, the non-Gaussian contributions can be read off from a Taylor 
expansion of the halo mass function.
For example, assuming that the Gaussian mass function of the tracers takes the form 
$\bar{n}(\delta_c,\sigma_0)$ and considering the constant-$\fnl\phi^2$ model for simplicity, 
the change induced by a long-wavelength perturbation 
$(\delta_l,\phi_l)$ is
\begin{equation}
\bar{n}\left[\delta_c-\delta_l,\sigma_0 (1+2\fnl\phi_l),S_3,\cdots\right] \approx 
\bar{n}(\delta_c,\sigma_0)
-\frac{\partial\bar{n}}{\partial\delta_c}\delta_l
+2\fnl\sigma_0\frac{\partial\bar{n}}{\partial\sigma_0}\phi_l 
+\cdots
\end{equation}
Therefore, the overabundance of tracers is given by
\begin{equation}
\delta_{\rm h}\approx 
-\frac{1}{\bar{n}}\frac{\partial\bar{n}}{\partial\delta_c}\delta_l
+2\fnl\left(\frac{\sigma_0}{\bar{n}}\frac{\partial\bar{n}}{\partial\sigma_0}\right)\phi_l 
+\cdots \equiv b_1 \delta_l+ 2\fnl b_1^{\rm NG} \phi_l +\cdots
\end{equation}
The non-Gaussian contribution to the first-order bias thus is proportional to the first 
derivative of average number density $\bar{n}$ of halos with respect to $\sigma_0$. 
However, while $b_1^{\rm NG}=b_1$ for a universal mass function, this will not hold in 
general. Is this simple intuitive picture also valid for the discrete peaks considered here?

\subsection{Application to discrete density peaks}

To answer this question, we shall first apply the peak-background split to the peak number 
density $\bnpk$, which is the ensemble average of (\ref{eq:npeak}). This was computed
in \cite{bardeen/etal:1986} as
\begin{equation}
\bnpk=\frac{1}{(2\pi)^2R_\star^3} G_0^{(1)}\!(\nu_c,\gamma_1\nu_c) e^{-\nu_c^2/2} \;.
\end{equation}
Interestingly, $\bnpk$ is {\em not} a universal function as it depends distinctly on $\delta_c$ 
and $R_s$. In fact, we can also assume that it depends on $\sigma_0$, $\sigma_1$ and 
$\sigma_2$ through the parameters $\gamma_1$ and $R_\star$, the significance $\nu_c$ and 
the normalized field $u$ (which is integrated over in $G_0^{(0)}$). 
Hence, we can write $\bnpk\equiv \bnpk(\delta_c,\{\sigma_i\})$, with $i=0,1,2$.
Therefore, the above discussion suggests that we take derivatives of $\bnpk$ with respect to the 
spectral moments $\sigma_i$. We begin with the derivative of $\bnpk$ with respect to $\sigma_0$ 
and easily obtain 
\begin{align}
\frac{\partial\bnpk}{\partial\sigma_0} &= 
\frac{\partial\bnpk}{\partial\nu_c}\frac{\partial\nu_c}{\partial\sigma_0}
+\frac{\partial\bnpk}{\partial\gamma_1}\frac{\partial\gamma_1}{\partial\sigma_0} 
= \frac{\bnpk}{\sigma_0}\Bigl(\sigma_0^2 b_{20}+1\Bigr) \;.
\end{align}
Similarly, the derivative of $\bnpk$ relative to the spectral moment $\sigma_1$ is 
\begin{align}
\frac{\partial\bnpk}{\partial\sigma_1} &= \frac{\partial\bnpk}{\partial R_\star}
\frac{\partial R_\star}{\partial\sigma_1} 
+\frac{\partial\bnpk}{\partial\gamma_1}\frac{\partial\gamma_1}{\partial\sigma_1} 
= 2\sigma_1\bnpk\Bigl(\chi_{10}+ b_{11}\Bigr) \;.
\end{align}
Finally, the derivative of $\bnpk$ with respect to $\sigma_2$ yields
\begin{align}
\frac{\partial\bnpk}{\partial\sigma_2} &=
\frac{1}{V_\star}\int_0^\infty\!\!du \,
\Biggl[f(u) \left(\frac{u}{\sigma_2}\right)
\left(\frac{u-\gamma_1\nu_c}{1-\gamma_1^2}\right)+\frac{\partial f(u)}{\partial\sigma_2}\Biggr]
\frac{e^{-\left(u-\gamma_1\nu_c\right)^2/\left[2\left(1-\gamma_1^2\right)\right]}}
{\sqrt{2\pi\left(1-\gamma_1^2\right)}}\, \frac{e^{-\nu_c^2/2}}{\sqrt{2\pi}} \nonumber \\
& \quad
-\frac{1}{\sigma_2} \bnpk 
+\frac{\partial\bnpk}{\partial R_\star}\frac{\partial R_\star}{\partial\sigma_2}
+\frac{\partial\bnpk}{\partial\gamma_1}\frac{\partial\gamma_1}{\partial\sigma_2} \;.
\end{align}
Here, $V_\star=(2\pi)^{3/2} R_\star^3$ and the term $-\bnpk/\sigma_2$ arises upon taking 
the derivative of the measure $du$ with respect to $\sigma_2$. The derivative of
$f(u)\equiv f(u,\alpha=1)$ produces a factor of $-8\bnpk/\sigma_2$, to which the measure 
$dv dw$ contributes $-2\bnpk/\sigma_2$ and the shape factor $F(u,v,w)$ (which is a 
homogeneous function of degree $-6$ in $\sigma_2$) the remaining $-6\bnpk/\sigma_2$. 
There is an additional contribution that originates from the dependence of $3v^2+w^2$ on 
$\sigma_2$ in the argument of the exponential. It can be written down as 
$-(2/\sigma_2)\partial_\alpha f(u,\alpha)\lvert_{\alpha=1}$. Adding up all the terms, we 
find
\begin{multline}
\frac{\partial\bnpk}{\partial\sigma_2} =
\frac{1}{\sigma_2}\left(\frac{\ov{u^2}-\gamma_1\nu_c\ov{u}}{1-\gamma_1^2}\right)\bnpk
-\frac{9}{\sigma_2}\bnpk -\frac{2}{\sigma_2}\partial_\alpha\log 
G_0^{(\alpha)}\!(\gamma_1,\gamma_1\nu_c)\Bigr\lvert_{\alpha=1} \bnpk \\
+\frac{\partial\bnpk}{\partial R_\star}\frac{\partial R_\star}{\partial\sigma_2}
+\frac{\partial\bnpk}{\partial\gamma_1}\frac{\partial\gamma_1}{\partial\sigma_2}
= \sigma_2 \bnpk\Bigl(2\chi_{01}+b_{02}\Bigr) \;.
\end{multline}
The logarithmic derivatives $(\bnpk/\sigma_i)\partial\bnpk/\partial\sigma_i$ thus are
\begin{equation}
\frac{\partial\log\bnpk}{\partial\log\sigma_0} = \sigma_0^2 b_{20}+1\, , \qquad
\frac{\partial\log\bnpk}{\partial\log\sigma_1} = 
2\sigma_1^2 \chi_{10}+2\sigma_1^2 b_{11}\, ,\qquad
\frac{\partial\log\bnpk}{\partial\log\sigma_2} = 2\sigma_2^2 \chi_{01}+\sigma_2^2 b_{02} \;.
\end{equation}
On multiplying the logarithmic derivatives by $\sigma_{\alpha+i}^2/\sigma_i^2$ and 
summing the resulting contributions, we arrive at 
\begin{equation}
\sum_{i=0}^2 \frac{\partial\log\bnpk}{\partial\log\sigma_i}
\left(\frac{\sigma_{\alpha+i}}{\sigma_i}\right)^2 = 
\sigma_\alpha^2 b_{20} + 2\sigma_{\alpha+1}^2 b_{11} + \sigma_{\alpha+2}^2 b_{02}
+2\sigma_{\alpha+1}^2 \chi_{10}+2\sigma_{\alpha+2}^2\chi_{01} 
+\frac{\sigma_\alpha^2}{\sigma_0^2} \, .
\label{eq:pbs_ampli}
\end{equation}
This is precisely the amplitude of the non-Gaussian bias correction found from the 
computation of the peak 2-point correlation, except for an additional factor of 
$\sigma_\alpha^2/\sigma_0^2$.

\subsection{From peaks to dark matter halos}

To explain the origin of this factor, we note that the differential number density of dark 
matter halos of mass $M$ per unit comoving volume is generically expressed as
\begin{equation}
\bar{n}_{\rm h}(M) = \frac{\bar{\rho}}{M}f(\nu_c)\frac{d\nu_c}{dM}=
\frac{\bar{\rho}}{M^2} \nu_c f(\nu_c) \frac{d\log\nu_c}{d\log M} \;,
\end{equation}
where the multiplicity function $f(\nu_c)$ encodes information about halo biasing. In the 
present calculation, the quantity $V\bnpk(\delta_c,\{\sigma_i\})$, where $V=M/\bar{\rho}$ 
is the Lagrangian volume associated with the filter ($M\propto R_s^3$), plays the role of 
a multiplicity function. 
Therefore, we can write the halo mass function associated with the peak number density 
(\ref{eq:npeak}) as
\begin{equation}
\bar{n}_{\rm h}(M)=\frac{\nu_c}{M}\,\bnpk(\delta_c,\{\sigma_i\})\frac{d\log\nu_c}{d\log M} \, .
\label{eq:nh_bbks}
\end{equation}
The logarithmic derivative of $\bar{n}_{\rm h}(M)$ with respect to $\sigma_i$ will be identical 
to that of $\bnpk$ except that, for $\sigma_0$, there will be an additional factor of $-1$ 
owing to the multiplicative factor of $\nu_c$ in the expression of the halo mass function. 
This factor precisely cancels the extra term of $\sigma_\alpha^2/\sigma_0^2$ in 
(\ref{eq:pbs_ampli}). Therefore, we find
\begin{equation}
\sum_{i=0}^2\frac{\partial\log\bar{n}_{\rm h}}{\partial\log\sigma_i}
\left(\frac{\sigma_{\alpha+i}}{\sigma_\alpha}\right)^2 =
\sigma_\alpha^2 b_{20} + 2\sigma_{\alpha+1}^2 b_{11} + \sigma_{\alpha+2}^2 b_{02}
+2\sigma_{\alpha+1}^2 \chi_{10}+2\sigma_{\alpha+2}^2\chi_{01} \, ,
\end{equation}
which is exactly the amplitude obtained from the calculation of the peak 2-point correlation.
The non-Gaussian contribution to the linear peak bias thus is
\begin{equation}
\Delta c_1(k) = 2A\frac{\fnl(k_p)}{k_p^{2\alpha_1}}
\Biggl[\sum_{i=0}^2\frac{\partial\log\bar{n}_{\rm h}}{\partial\log\sigma_i}
\left(\frac{\sigma_{\alpha+i}}{\sigma_\alpha}\right)^2\Biggr] k^{-2\alpha_2}\calM^{-1}(k)
\, .
\end{equation}
The physical interpretation is straightforward: in the presence of a primordial 3-point 
function, a long-wavelength background perturbation of wavenumber $k$ rescales the 
amplitude of the power spectrum $P_s(q)$ of the smoothed density field in a scale-dependent 
manner \cite{slosar/etal:2008,schmidt/kamionkowski:2010,desjacques/etal:2011a},
\begin{equation}
P_s(q) ~\to~ \left(1+2\frac{\delta\sigma_8}{\sigma_8}\right)P_s(q) \approx 
\left[1+2\epsilon\biggl(\frac{q}{k_p}\biggr)^{2\alpha_1}\biggl(\frac{q}{k}\biggr)^{2\alpha_2}\right] 
P_s(q) \;,
\label{eq:pstransform}
\end{equation}
where $\epsilon$ will be determined shortly. Note that this approximation is valid in the 
limit $k\ll q$ solely.
In the particular case of the constant-$\fnl\phi^2$ model, all the spectral moments are 
rescaled analogously, $\sigma_i\to (1+\epsilon)\sigma_i$, so that the parameters $\gamma_1$ 
and $R_\star$ remain unchanged. 
In general however, the transformation $\sigma_i\to \sigma_i+\delta\sigma_i$ will be 
scale-dependent. We can write the derivative of $\bar{n}_{\rm h}$ with respect to the 
normalisation amplitude as
\begin{equation}
\frac{\partial\log\bar{n}_{\rm h}}{\partial\sigma_8}\delta\sigma_8
\equiv\sum_{i=0}^2 \frac{\partial\log\bar{n}_{\rm h}}{\partial\log\sigma_i}
\left(\frac{\delta\sigma_i}{\sigma_i}\right)
=\epsilon k_p^{-2\alpha_1}
\sum_{i=0}^2 \frac{\partial\log\bar{n}_{\rm h}}{\partial\log\sigma_i}
\left(\frac{\sigma_{\alpha+i}}{\sigma_i}\right)^2 \;.
\end{equation}
On setting $\epsilon \equiv 2\fnl(k_p)\calM^{-1}(k)$, we recover the full non-Gaussian 
$k$-dependent correction to the linear halo bias.

\section{Extension to excursion set peaks}
\label{sec:firstcrossing}

The differential peak number density $\bnpk$ cannot really be 
interpreted as a multiplicity function because it is defined for a fixed smoothing scale 
$R_s$, whereas one should allow $R_s$ to vary while $\delta=\delta_c$ is kept fixed. 
This is the reason why we have not yet recovered the strong mass-dependent correction 
found by \cite{desjacques/etal:2011a}.
Furthermore, the linear bias factor $b_{10}$ is not equal to the Gaussian peak-background 
split bias $-d\ln\bar{n}_{\rm h}/d\delta_c$ inferred from a uniform shift 
$\delta_c\to\delta_c+\epsilon_1$ applied to the halo mass function (\ref{eq:nh_bbks}).

Therefore, this suggests that we consider the improved model of 
\cite{appel/jones:1990,paranjape/sheth:2012}, in which 
peaks on a given smoothing scale contribute to the multiplicity function only if the 
conditions $\delta(R_s)>\delta_c$ and $\delta(R_s+\Delta R_s)<\delta_c$ are satisfied. 
In other words, the density must upcross the threshold for collapse on the smoothing 
scale $R_s$. 
Because the trajectory described by the Gaussian or tophat filtered $\delta(R_s)$ as a 
function of $R_s$ is strongly correlated when $\sigma_0(R_s)\lesssim 1$, this almost 
certainly implies that $\delta(R_s)$ upcrosses the threshold for the first time at 
$R=R_s$ (see, e.g. \cite{musso/paranjape:2012} for a detailed discussion).

\subsection{Effective local bias expansion}

To include the first crossing condition into the calculation of the non-Gaussian bias, 
we must introduce a new variable: $\mu\equiv -d\delta_s/dR_s = -\delta_s'$ 
(this notation is borrowed from \cite{appel/jones:1990} who considered the particular 
case of a Gaussian filter). 
The number density of density peaks for which the first upcrossing occurs on the 
filtering scale $R_s$ is
\begin{align}
\bar{n}_{\rm UC}(\nu_c,R_s)\Delta R_s &= 
\frac{3^{3/2}}{R_\star^3}
\int\!\! d^6\pmb\zeta \int\!\!d^3\pmb\eta \int_{-\infty}^0\!\!d\delta'
\int_{\delta_c}^{\delta_c-\delta'\Delta R_s}\!\! \frac{d\delta}{\sigma_0}\,
\left|\det\pmb{\zeta}\right| 
\delta_D\!\left(\pmb{\eta}\right) \theta_H\!\left(\lambda_3\right)
P_1(\mathbi{w}) 
\nonumber \\
&= \int\!\!d^6\pmb\zeta \int\!\!d^3\pmb\eta \int\!\!d\nu \int_0^\infty\!\!d\mu\,
\frac{\mu}{\sigma_0} \npk(\mathbi{y})
\, P_1(\mathbi{w}) \Delta R_s \;,
\end{align}
where $\mathbi{w}$ is the 11-dimensional vector of variables 
$\mathbi{w}=(\mu,\nu,\pmb\eta,\pmb\zeta)=(\mu,\mathbi{y})$. 
Therefore, we can write the excursion set peaks multiplicity function as 
\begin{equation}
\fesp(\nu_c,R_s)=\frac{M}{\bar{\rho}}\,\bar{n}_{\rm UC}(\nu_c,R_s)\frac{dR_s}{d\nu_c}
= -\frac{V}{\nu_c\sigma_0'}
\int\!\!d^{11}\mathbi{w}\,\mu\,\theta_H(\mu)\,\npk(\mathbi{y}) P_1(\mathbi{w}) \;,
\end{equation}
where it is understood that $R_s$ is allowed to vary while the density threshold $\delta_c$ 
is kept fixed. We can thus think of excursion set peaks as arising from the discrete number 
density
\begin{equation}
\label{eq:nesp}
\esp(\mathbi{w}) = -\frac{\mu}{\nu_c\sigma_0'} \theta_H(\mu)\, \npk(\mathbi{y}) \;.
\end{equation}
Our calculation is valid for any smoothing kernel. In the special case of Gaussian filtering,
$\mu=R_s \sigma_2 u$ and $\sigma_0'=-R_s\sigma_1^2/\sigma_0$, and we recover the prefactor of 
$u/(\gamma_1\nu_c)$ obtained by \cite{paranjape/sheth:2012}
[the step function $\theta_H(\mu)=\theta_H(u)$ then becomes redundant with $\theta_H(\lambda_3)$]. 
For practical purposes, the excursion set peaks multiplicity function can be computed from
\begin{equation}
\fesp(\nu_c,R_s)=-\frac{V}{V_\star}
\,\frac{G_0^{(1)}\!(\gamma_1,\gamma_{u\mu},\nu_c)}{\nu_c\sigma_0'}\, 
\frac{e^{-\nu_c^2/2}}{\sqrt{2\pi}}\;,
\end{equation}
where $G_n^{(\alpha)}$ is generalized to
\begin{equation}
G_n^{(\alpha)}\!(\gamma_1,\gamma_{u\mu},\nu_c)\equiv 
\int_0^\infty\!\!d\mu\,\mu{\cal N}(\mu)
\int_0^\infty\!\!du\,u^n f(u,\alpha) {\cal N}(u|\nu_c,\mu) \;,
\end{equation}
with $\gamma_{u\mu}$ given shortly.
Note that the multiplicative factor of $f(u,\alpha)$ remains unchanged because the variable 
$\zeta^2$ is uncorrelated with $(\nu,u,\mu)$. 

Following \cite{desjacques:2012}, rotational invariance implies that the 1-point probability 
density $P_1(\mathbi{w})$ be written as
\begin{equation}
\label{ESP1pointP}
P_1(\mathbi{w}) d^{11}\mathbi{w} = {\cal N}(\nu,u,\mu) d\nu du d\mu \times \chi_3^2(3\eta^2) d(3\eta^2)
\times \chi_5^2(5\zeta^2) d(5\zeta^2) \;, 
\end{equation}
where ${\cal N}(\nu,u,\mu)$ is a trivariate normal distribution and $\chi_k^2(x)$ is a 
$\chi^2$-distribution. The cross-correlations between the variables $\nu$, $u$ and $\mu$ are
\begin{equation}
\left\la\nu u\right\ra = \gamma_1\;, \qquad
\left\la\nu\mu\right\ra = -\sigma_0' \equiv \gamma_{\nu\mu}\;, \qquad
\left\la u\mu\right\ra = -\frac{\sigma_1}{\sigma_2}\sigma_1' \equiv \gamma_{u\mu} \;,
\end{equation}
whereas $\la\nu^2\ra=\la u^2\ra =1$ and $\la\mu^2\ra=\la(\delta')^2\ra\equiv\Delta_0^2$. Owing
to the new scalar variable $\mu$, the effective local bias relation (\ref{eq:dpk}) must be 
generalized to
\begin{align}
\dpk(\mathbi{x}) &= \sigma_0 b_{100}\nu(\mathbi{x})+\sigma_2 b_{010} u(\mathbi{x}) 
+ b_{001}\mu(\mathbi{x}) 
\nonumber \\
&\quad + \frac{1}{2}\sigma_0^2 b_{200}\nu^2(\mathbi{x})
+\sigma_0\sigma_2 b_{110}\nu(\mathbi{x}) u(\mathbi{x})
+\frac{1}{2}\sigma_2^2 b_{020} u^2(\mathbi{x})+\sigma_1^2\chi_{10}\eta^2(\mathbi{x}) 
+\sigma_2^2\chi_{01}\zeta^2(\mathbi{x}) \nonumber \\
&\quad + \frac{1}{2}b_{002}\mu^2(\mathbi{x}) + \sigma_0 b_{101}\nu(\mathbi{x})\mu(\mathbi{x}) 
+\sigma_2 b_{011}u(\mathbi{x})\mu(\mathbi{x}) + \cdots
\label{eq:newdpk}
\end{align}
The bias parameters $b_{ijk}$ are defined as ensemble averages over the trivariate Hermite 
polynomials constructed from ${\cal N}(\nu,u,\mu)$,
\begin{equation}
\sigma_0^i \sigma_2^j b_{ijk} = \frac{1}{\besp}\int\!\!d^{11}\mathbi{w}\,\esp(\mathbi{w}) 
H_{ijk}(\nu,u,\mu) P_1(\mathbi{w}) \;,
\end{equation}
where $\besp\equiv \fesp/V$ is the number density of excursion set peaks. There is no factor of 
$\Delta_0^k$ in the left-hand side since $\mu$, unlike $\nu$ and $u$, is not normalized to
have unit variance.

The excursion set peaks multiplicity function scales as $\fesp\propto \exp(-\nu_c^2/2)/\nu_c$, 
which should be compared to $f(\nu_c)=\exp(-\nu_c^2/2)$ in the Press-Schechter formalism. 
The presence of a multiplicative factor of $\nu_c^{-1}$ guarantees that the Gaussian bias 
factors $b_{k00}$, which involve derivatives of $P_1$ with respect to the peak height, agree 
exactly  with those obtained from a peak-background split applied to the mass function 
$\bar{n}_{\rm h}(M)\propto \nu_c \fesp(\nu_c,R_s)$ of dark matter halos.

\subsection{Non-Gaussian bias}

In light of the results derived in Section~\ref{sec:pkngbias}, a quick calculation shows that 
the non-Gaussian correction to the power spectrum of excursion set peaks is
\begin{align}
\Delta P_{\rm pk}(k) &= 4A \frac{\fnl(k_p)}{k_p^{2\alpha_1}} k^{-2\alpha_2}\calM_s(k) P_\phi(k)
\left(b_{100}+ b_{010}k^2+\cdots\right) 
\nonumber \\
& \quad \times \biggl[\sigma_\alpha^2 b_{200} + 2\sigma_{\alpha+1}^2 b_{110} 
+ \sigma_{\alpha+2}^2 b_{020} + 2\sigma_{\alpha+1}^2 \chi_{10}+2\sigma_{\alpha+2}^2\chi_{01}
\nonumber \\
&\qquad + \Delta_\alpha^2 b_{002}- \bigl(\sigma_{\alpha}^2\bigr)'\, b_{101}
-\bigl(\sigma_{\alpha+1}^2\bigr)'\, b_{011}\biggr] \;.
\label{eq:dpkesp}
\end{align}
The symbol $\Delta_\alpha^2$ is similar to the spectral moment $\sigma_\alpha^2$ defined
above, except that the density field $\delta$ is replaced by its derivative $\mu$ with respect to 
the filtering scale:
\begin{equation}
\Delta_\alpha^2 \equiv \frac{1}{2\pi^2}\int_0^\infty\!\! dk\, k^{2(\alpha+1)} 
\bigl[\calM_s'(k)\bigr]^2 P_\phi(k) \;.
\end{equation}
We shall now proceed analogously to Section~\ref{sec:pbs} and compare the amplitude of the 
non-Gaussian correction (\ref{eq:dpkesp}) with logarithmic derivatives of the halo mass 
function,
\begin{equation}
\label{eq:nhalo}
\bar{n}_{\rm h}(M) = \frac{\bar{\rho}}{M^2} \nu_c \fesp(\nu_c,R_s)\frac{d\log\nu_c}{d\log M} \;.
\end{equation}
constructed from the multiplicity function of excursion set peaks. The subtlety resides in
handling the derivatives with respect to the smoothing radius $R_s$. 

\begin{figure}
\center
\resizebox{0.80\textwidth}{!}{\includegraphics{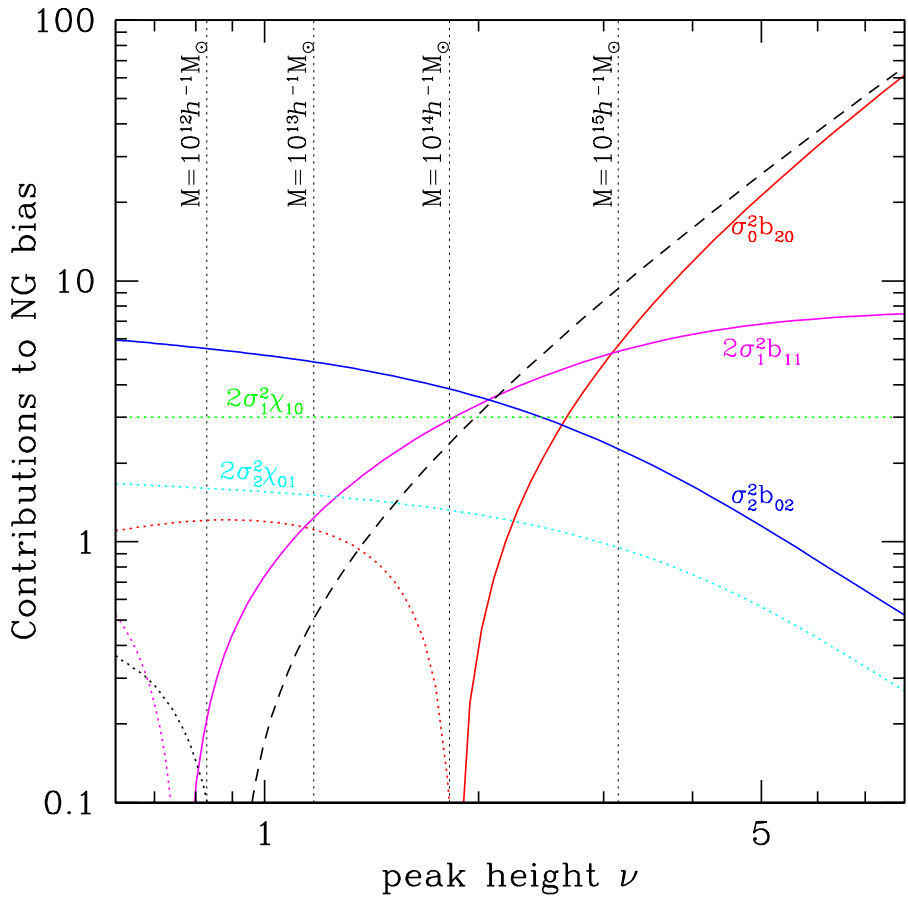}}
\caption{Contributions of the second-order Lagrangian peak bias factors to the 
amplitude of the non-Gaussian bias in the local $\fnl\phi^2$ model (for which $\alpha=0$). 
The various curves have been derived from the excursion set peaks mass function of 
\cite{paranjape/sheth/desjacques:2012} (see text for details). 
All the second-order bias parameters are multiplied by the appropriate factors of $\sigma_i$ 
so that all terms are dimensionless. The dashed curve represents the sum of all contributions.
Vertical lines mark the peak significance at which the corresponding halo mass is in the range 
$M=10^{12} - 10^{15}\hmsun$.}
\label{fig:bias}
\end{figure}

To get a sense of the importance of the various second-order bias terms in Eq.(\ref{eq:dpkesp}),
we plot in Fig.\ref{fig:bias} their relative contribution to the amplitude of the non-Gaussian 
bias in the local $\fnl$ model. We consider the excursion set peak mass function proposed by 
\cite{paranjape/sheth/desjacques:2012}. 
However, while their approach actually implies $\mu\ne R_s\sigma_2 u$ (since $\mu$ is smoothed 
with a tophat while $u$ is smoothed with a Gaussian), we assume that the equality holds for 
simplicity and, thus, ignore terms involving $\mu(\vx)$ in the effective peak bias expansion. 
In addition, while we adopt the same mean moving barrier $B(\sigma_0)=\delta_c+0.43\sigma_0$ 
as \cite{paranjape/sheth/desjacques:2012}, we ignore the scatter in collapse thresholds since
all this is for illustrative purposes only.
The curves in Fig.\ref{fig:bias} thus only represent the contributions $\sigma_0^2b_{20}$, 
$2\sigma_1^2 b_{11}$, $\sigma_2^2 b_{02}$, $2\sigma_1^2\chi_{10}$ and $2\sigma_2^2\chi_{01}$,
and are labelled accordingly. The dashed curve is the sum of all these terms.
For $\nu\lesssim 1$, the relative contribution of $\sigma_2^2 b_{20}$ (which arise from 
$u^2(\vx)$) dominates the amplitude of the non-Gaussian bias, whereas those of 
$2\sigma_1^2\chi_{10}$ and $2\sigma_2^2\chi_{01}$ (induced by $\eta^2(\vx)$ and $\zeta^2(\vx)$,
respectively) are of negative sign.

\subsection{Consistency with peak-background split}

In the presence of a long-wavelength background perturbation, the power spectrum $P_s(q)$
of the smoothed density field is rescaled according to (\ref{eq:pstransform}) so that,
in the large scale limit $k\ll 1$ considered in this work, the spectral moments $\sigma_i$ 
and $\sigma_i'$ transform as
\begin{align}
\sigma_i ~&\to~ \sigma_i\left(1+\frac{\delta\sigma_i}{\sigma_i}\right)
\equiv\sigma_i\left(1+\epsilon \frac{\sigma_{\alpha+i}^2}{\sigma_i^2}\right) 
\nonumber \\
\sigma_i' ~&\to~ \sigma_i'\left(1+\frac{\delta\sigma_i'}{\sigma_i'}\right)
\equiv \sigma_i'\Biggl[1+\epsilon\frac{\sigma_{\alpha+i}^2}{\sigma_i^2}
+2\epsilon\frac{\sigma_{\alpha+i}^2}{\sigma_i^2}\left(\frac{\partial\log\sigma_{\alpha+i}}
{\partial\log\sigma_i}-1\right)\Biggr] \;,
\label{eq:dsigmas}
\end{align}
where $\epsilon \equiv 2\fnl(k_p)\calM^{-1}(k)$. 
Hence, both $\sigma_i$ and $\sigma_i'$ equally contribute a term $\epsilon (\sigma_{\alpha+i}/
\sigma_i)^2$, whereas $\sigma_i'$ solely induces a correction proportional to 
$\partial\log\sigma_{\alpha+i}/\partial\log\sigma_i$. Therefore, to compute the 
correction proportional to, e.g. $(\sigma_{\alpha}/\sigma_0)^2$ from peak-background split, 
we must add the logarithmic derivatives of $n_{\rm h}(M)$ with respect to both $\sigma_0$ and 
$\sigma_0'$. On reformulating the halo mass function in the more convenient form
\begin{equation}
\bar{n}_{\rm h}(M) = \frac{\bar{\rho}}{M}\left[-\nu_c \fesp(\nu_c,R_s)\frac{\sigma_0'}{\sigma_0}\right]
\frac{dR}{dM}
\label{eq:nhfnu}
\end{equation}
and taking into account the dependence of the excursion set peaks multiplicity function on 
$\sigma_0'$, we find
\begin{align}
\frac{\partial\log\bar{n}_{\rm h}}{\partial\log\sigma_0}
+\frac{\partial\log\bar{n}_{\rm h}}{\partial\log\sigma_0'} &= 
\biggl(
-\frac{\partial\log\fesp}{\partial\log\nu_c}-\frac{\partial\log\fesp}{\partial\log\gamma_1}-2\biggr)
+\frac{\partial\log\fesp}{\partial\log\gamma_{\nu\mu}}\nonumber \\
&=\biggl(-\left\la\frac{\partial\log P_1}{\partial\log\nu}\right\ra
-\frac{\partial\log\fesp}{\partial\log\gamma_1}-1\biggr)
+\frac{\partial\log\fesp}{\partial\log\gamma_{\nu\mu}} \nonumber \\
&\equiv \sigma_0^2 b_{200}+2\gamma_{\nu\mu}\sigma_0 b_{101} \;,
\end{align}
where $\left\la X\right\ra \equiv \int X \esp P_1/\besp$.
The last expression involves second derivatives of the 1-point probability density $P_1(\mathbi{w})$ 
with respect to the variables $\nu$ and $\mu$. In particular, 
\begin{equation}
\frac{\partial\log\bar{n}_{\rm h}}{\partial\log\sigma_0'}=
\frac{\partial\log\fesp}{\partial\log\sigma_0'}+1=
\frac{\partial\log\fesp}{\partial\log\gamma_{\nu\mu}}=\gamma_{\nu\mu}\sigma_0 b_{101} \;.
\end{equation}
These relations can now be substituted into the terms proportional to $\sigma_\alpha^2$ and 
$(\sigma_\alpha^2)'$ in the square brackets of (\ref{eq:dpkesp}),
\begin{align}
\sigma_{\alpha}^2 b_{200}-(\sigma_\alpha^2)' b_{101}
&= \bigl(\sigma_0^2 b_{200}+2\gamma_{\nu\mu}\sigma_0 b_{101}\bigr)
\left(\frac{\sigma_\alpha^2}{\sigma_0^2}\right)
+\gamma_{\nu\mu}\sigma_0 b_{101}\Biggl[2\frac{\sigma_\alpha^2}{\sigma_0^2}
\left(\frac{\partial\log\sigma_\alpha}{\partial\log\sigma_0}-1\right)\Biggr]\nonumber \\
&\equiv \frac{\partial\log\bar{n}_{\rm h}}{\partial\log\sigma_0} 
\left(\frac{\delta\sigma_0}{\sigma_0}\right)
+\frac{\partial\log\bar{n}_{\rm h}}{\partial\log\sigma_0'} 
\left(\frac{\delta\sigma_0'}{\sigma_0'}\right) \;.
\end{align}
For $\sigma_1$ and $\sigma_1'$, we must consider derivatives of the halo mass function with 
respect to $R_\star$, $\gamma_1$ and $\gamma_{u\mu}$. A straightfoward calculation leads to
\begin{equation}
\frac{\partial\log\bar{n}_{\rm h}}{\partial\log\sigma_1'} =
\frac{\partial\log\fesp}{\partial\log\sigma_1'}= 
\frac{\partial\log\fesp}{\partial\log\gamma_{u\mu}}=
\sigma_2 \gamma_{u\mu}b_{011} \;,
\end{equation}
and
\begin{align}
\frac{\partial\log\bar{n}_{\rm h}}{\partial\log\sigma_1}+
\frac{\partial\log\bar{n}_{\rm h}}{\partial\log\sigma_1'} &=
2\frac{\partial\log\fesp}{\partial\log\gamma_1}+\frac{\partial\log\fesp}{\partial\log R_\star}
+2\frac{\partial\log\fesp}{\partial\log\gamma_{u\nu}}
= 2\sigma_1^2 \bigl(b_{110}+\chi_{10}\bigr)+\sigma_2\gamma_{u\mu}b_{011} \;.
\end{align}
Therefore,
\begin{align}
2 \sigma_{\alpha+1}^2 \bigl(b_{110}&+\chi_{10}\bigr)-(\sigma_{\alpha+1}^2)' b_{011} \nonumber\\
&= \Bigl[2\sigma_1^2\bigl(b_{110}+\chi_{10}\bigr)+2\sigma_2\gamma_{u\mu}b_{011}\Bigr]
\left(\frac{\sigma_{\alpha+1}^2}{\sigma_1^2}\right)
+\sigma_2\gamma_{u\mu} b_{011}\left[2\frac{\sigma_{\alpha+1}^2}{\sigma_1^2}
\left(\frac{\partial\log\sigma_{\alpha+1}}{\partial\log\sigma_1}-1\right)\right] \nonumber \\
&\equiv \frac{\partial\log\bar{n}_{\rm h}}{\partial\log\sigma_1} 
\left(\frac{\delta\sigma_1}{\sigma_1}\right)
+\frac{\partial\log\bar{n}_{\rm h}}{\partial\log\sigma_1'} 
\left(\frac{\delta\sigma_1'}{\sigma_1'}\right) \;.
\end{align}
For $\sigma_2$, the calculation is somewhat more intricate. On the one hand, taking the 
derivative of the halo mass function with respect to $\sigma_2$, we arrive at
\begin{align}
\frac{\partial\bar{n}_{\rm h}}{\partial\sigma_2}&= 
\frac{\partial\fesp}{\partial\gamma_1}\frac{\partial\gamma_1}{\partial\sigma_2}
+\frac{\partial\fesp}{\partial R_\star}\frac{\partial R_\star}{\partial\sigma_2}
+\frac{\partial\fesp}{\partial\gamma_{u\mu}}\frac{\partial\gamma_{u\mu}}{\partial\sigma_2}
\nonumber \\
&\quad -\frac{9}{\sigma_2}\fesp
+\frac{2}{\sigma_2}\left(\frac{V}{V_\star}\right)\bigl(\nu_c\sigma_0'\bigr)^{-1}
\partial_\alpha G_0^{(\alpha)}\!(\gamma_1,\gamma_{u\mu},\nu_c) \frac{e^{-\nu_c^2/2}}{\sqrt{2\pi}}
+\left\la\frac{\partial P_1}{\partial u}\right\ra \frac{\partial u}{\partial\sigma_2} \;.
\label{eq:step1}
\end{align}
The term $-9\fesp/\sigma_2$ includes the contribution from $F(u,w,v)$ (a factor of $-6$) and 
from the measure $du dv dw$ (a factor of $-3$). On the other hand, the terms proportional to 
$\sigma_{\alpha+2}$ in the non-Gaussian bias correction $\Delta c_1(k)$ are
\begin{equation}
\label{eq:ds1}
\sigma_{\alpha+2}^2 b_{020}+2\sigma_{\alpha+2}^2\chi_{01} =
\Bigl(\sigma_2^2 b_{020}+2\sigma_2^2\chi_{01}\Bigr) \frac{\sigma_{\alpha+2}^2}{\sigma_2^2} \;.
\end{equation}
We can easily convince ourselves that
\begin{equation}
\sigma_2^2 b_{020} = -\frac{\partial\log\fesp}{\partial\log\gamma_1}
-\frac{\partial\log\fesp}{\partial\log\gamma_{u\mu}}-\frac{\sigma_2}{\fesp}
\left\la\frac{\partial P_1}{\partial u}\right\ra -1 \;.
\end{equation}
Therefore,
\begin{align}
\sigma_2^2 b_{020}+2\sigma_2^2\chi_{01} &=
-\frac{\partial\log\fesp}{\partial\log\gamma_1}
-\frac{\partial\log\fesp}{\partial\log\gamma_{u\mu}}-\frac{\sigma_2}{\fesp}
\left\la\frac{\partial P_1}{\partial u}\right\ra -1  
\nonumber \\
& \quad -\frac{\partial\log\fesp}{\partial\log R_\star} -8 
-2\partial_\alpha\log G_0^{(\alpha)}\!(\gamma_1,\gamma_{u\mu},\nu_c)\biggr\lvert_{\alpha=1} \;,
\label{eq:step2}
\end{align}
where, in the second line of the right-hand side, we wrote $-5=-8+3$ and used the fact that 
$-\partial\log\fesp/\partial\log R_\star = 3$. 
On multiplying (\ref{eq:step1}) by $(\sigma_2/\bar{n}_{\rm h})$ and comparing with 
(\ref{eq:step2}), we find that
\begin{equation}
\sigma_{\alpha+2}^2 b_{020}+2\sigma_{\alpha+2}\chi_{01} = 
\frac{\partial\log\bar{n}_{\rm h}}{\partial\log\sigma_2}
\left(\frac{\sigma_{\alpha+2}^2}{\sigma_2^2}\right)
\equiv \frac{\partial\log\bar{n}_{\rm h}}{\partial\log\sigma_2}
\left(\frac{\delta\sigma_2}{\sigma_2}\right) \;.
\end{equation}
Finally, the last contribution to be checked is that from $\Delta_\alpha^2$. The second-order
bias factors $b_{002}$ is proportional to the logarithmic derivative of $\fesp$ with respect to
$\Delta_0$, i.e.
\begin{equation}
\Delta_0^2\, b_{002} \equiv \frac{\partial\log\bar{n}_{\rm h}}{\partial\log\Delta_0} \;.
\end{equation} 
Therefore, observing that $\delta\Delta_0/\Delta_0=\Delta_\alpha^2/\Delta_0^2$, the square
brackets of (\ref{eq:dpkesp}) (i.e. the sum of second-order bias coefficients) can be expressed 
as
\begin{equation}
\frac{\partial\log\bar{n}_{\rm h}}{\partial\sigma_8}\delta\sigma_8 \equiv 
\sum_{i=0,1} \biggl[
\frac{\partial\log\bar{n}_{\rm h}}{\partial\log\sigma_i}
\left(\frac{\delta\sigma_i}{\sigma_i}\right)+
\frac{\partial\log\bar{n}_{\rm h}}{\partial\log\sigma_i'}
\left(\frac{\delta\sigma_i'}{\sigma_i'}\right)\biggr]
+\frac{\partial\log\bar{n}_{\rm h}}{\partial\log\sigma_2}
\left(\frac{\delta\sigma_2}{\sigma_2}\right)
+\frac{\partial\log\bar{n}_{\rm h}}{\partial\log\Delta_0}
\left(\frac{\delta\Delta_0}{\Delta_0}\right)
\label{eq:dbksigma8}
\end{equation}
where the change in the normalisation amplitude $\sigma_8$ is generally scale-dependent.  Note that 
the term $\delta\sigma_0'/\sigma_0'$ brings down the factor of 
$(\partial\log\sigma_\alpha/\partial\log\sigma_0-1)$ found by \cite{desjacques/etal:2011a}, 
which leads to pronounced effects when the primordial non-Gaussianity is not of the 
constant-$\fnl\phi^2$ form (see \cite{desjacques/etal:2011b}). 
In the excursion set peaks formalism, this correction appears from the requirement that $R_s$ (or $M$) 
varies while $\delta(\mathbi{x})=\delta_c$ is kept fixed.

\subsection{Non-Gaussian bias and non-universality of the mass function}

We emphasize again that the halo mass function derived from excursion set peaks is not universal. 
In this regards, \cite{scoccimarro/etal:2012} derived an expression for the non-Gaussian correction 
to the first-order bias which is valid for non-universal multiplicity functions of the form 
$f(\delta_c,\sigma_0)$. 
In the low-$k$ limit and for a primordial 3-point function, their result reduces to [omitting a 
factor of $2A\fnl(k_p)/\calM_s(k)k_p^{-2\alpha_1} k^{-2\alpha_2}$ for clarity]
\begin{multline}
\frac{\partial}{\partial M}\!\left[\frac{\sigma_\alpha^2}{\sigma_0^2}
\nu_c f(\delta_c,\sigma_0)\right]
\left[\frac{\nu_c}{\sigma_0}f(\delta_c,\sigma_0)\frac{d\sigma_0}{dM}\right]^{-1} 
\\ 
= \biggl\{2\left(\frac{\partial\log\sigma_\alpha}{\partial\log\sigma_0}-1\right)
+\frac{\partial}{\partial M}\log\bigl[\nu_c f(\delta_c,\sigma_0)\bigr]
\left(\frac{1}{\sigma_0}\frac{d\sigma_0}{dM}\right)^{-1}\biggr\}
\frac{\sigma_\alpha^2}{\sigma_0^2} \;.
\end{multline}
The main difference with our findings is the presence of a derivative with respect to the halo 
mass rather than the normalisation amplitude $\sigma_8$. Does their formula really differs from 
ours?

To answer this question, we substitute $\fesp$ into the above expression and write the derivative 
with respect to the halo mass as $(\partial/\partial M)=(\partial/\partial\sigma_0)(d\sigma_0/dM)+
(\partial/\partial\sigma_0')(d\sigma_0'/dM)+\cdots$. Using (\ref{eq:dsigmas}), we eventually 
arrive at 
\begin{align}
\frac{\partial}{\partial M}\!\biggl(\frac{\sigma_\alpha^2}{\sigma_0^2}
\nu_c & \fesp\biggr)
\left(\frac{\nu_c}{\sigma_0}\fesp\frac{d\sigma_0}{dM}\right)^{-1} \nonumber \\
&= \left(\frac{\partial\log\fesp}{\partial\log\sigma_0}+\frac{\partial\log\fesp}{\partial\log\sigma_0}
-1\right)\frac{\sigma_\alpha^2}{\sigma_0^2}
+2\left(\frac{\partial\log\fesp}{\partial\log\sigma_0'}+1\right)\frac{\sigma_\alpha^2}{\sigma_0^2}
\left(\frac{\partial\log\sigma_\alpha}{\partial\log\sigma_0}-1\right) \nonumber \\
&\quad 
+\left(\frac{\partial\log\fesp}{\partial\log\sigma_1}+\frac{\partial\log\fesp}{\partial\log\sigma_1'}
\right)\frac{\sigma_{\alpha+1}^2}{\sigma_1^2}
+2\left(\frac{\partial\log\fesp}{\partial\log\sigma_1'}\right)\frac{\sigma_{\alpha+1}^2}{\sigma_1^2}
\left(\frac{\partial\log\sigma_{\alpha+1}}{\partial\log\sigma_1}-1\right)  \nonumber \\
& \quad
+ \left(\frac{\partial\log\fesp}{\partial\log\sigma_2}\right)\frac{\sigma_{\alpha+2}^2}{\sigma_2^2}
+ \left(\frac{\partial\log\fesp}{\partial\log\Delta_0}\right)\frac{\Delta_\alpha^2}{\Delta_0^2} \;.
\label{eq:dbkmass}
\end{align} 
A comparison of the right-hand side with expressions derived above shows that it is exactly equal 
to (\ref{eq:dbksigma8}) [a rather unsurprising result given that smoothing kernels always
appear through $P_s(q)$]. 
Therefore, for excursion set peaks, the square bracket of (\ref{eq:dpkesp}), (\ref{eq:dbksigma8}) 
and (\ref{eq:dbkmass}) are all equivalent expressions for the amplitude of the non-Gaussian 
correction to the linear halo bias.

The consistency of the formalism considered here is now established. The non-Gaussian correction 
to the first-order peak bias is a weighted sum over the second-order peak bias factors. These
nicely combine into a logarithmic derivative of the halo mass function with respect to the normalisation
amplitude (e.g. $\sigma_8$), as expected from peak-background split. The first-crossing 
condition ensures that the correction found by \cite{desjacques/etal:2011a} be present and that 
the $k$-independent piece of the Gaussian peak bias factors satisfy the peak-background split 
relation $b_{k00}\equiv (-1)^k\bar{n}_{\rm h}^{-1} d^k\bar{n}_{\rm h}/d\delta_c^k$.

\section{Discussion of the results}
\label{sec:discussion}

We shall now draw connections with previous analytical works on the non-Gaussian bias, focusing 
on thresholded regions, and discuss why the standard Lagrangian local bias model fails at predicting 
the correct non-Gaussian bias.

\subsection{Non-Gaussian bias of  thresholded regions}

Matsubara \cite{matsubara:2012} worked out non-Gaussian bias corrections for rather generic 
Lagangian bias relations. On large scales, the leading order contribution to the non-Gaussian bias 
was found to be
\begin{equation}
\Delta b_1(k)\approx \frac{Q_2(k)}{2P_s(k)}=\frac{1}{2P_s(k)}
\int\!\!\frac{d^3\mathbi{q}}{(2\pi)^3}\,c_2(\mathbi{q},\mathbi{k}-\mathbi{q})
B_s(q,k,|\mathbi{k}-\mathbi{q}|)\;.
\end{equation}
In addition, the continuous number density $n(\mathbi{x},M)$ of halos of mass $M$ was defined in 
such a way that the actual halo mass function is recovered under spatial averaging. 
In the particular case of regions with overdensity equal to the critical thresholded $\delta_c$
(what we refer to hereafter as thresholded regions), one has
\begin{equation}
n(\mathbi{x},M)=-\frac{2\bar{\rho}}{M} 
\frac{\partial}{\partial M}\theta_H\bigl[\delta_s(\mathbi{x})-\delta_c\bigr] \;.
\end{equation}
It can be easily checked that $\left\la\theta_H[\delta_s(\mathbi{x})-\delta_c]\right\ra=P(M,\delta_c)$, 
the probability that $\delta_s(\mathbi{x})$ exceeds the critical threshold
for collapse. To extend the scope of his calculation, Matsubara assumed that $n(\mathbi{x},M)$ may be
different from $\partial_M\theta_H(\delta_s-\delta_c)$.
For universal multiplicity functions, the second-order renormalized bias parameter was shown
to take the form
\begin{equation}
\label{eq:c2k1k2}
c_2(\mathbi{k}_1,\mathbi{k}_2) = b_2(M) W(k_1 R_s) W(k_2 R_s)
+ \frac{1+\delta_c b_1(M)}{\delta_c^2} 
\frac{\partial}{\partial\log\sigma_0}\Bigl[W(k_1 R_s) W(k_2 R_s)\Bigr] \;,
\end{equation}
where $b_1$ and $b_2$ are the usual Gaussian peak-background split bias parameters.
Note that factors of $\partial/\partial\log M$ appear at all orders. As we will see shortly, they 
are equivalent to the $\mu^n(\mathbi{x})$ present in the effective local bias expansion (\ref{eq:newdpk}). 
As a result, the function $Q_2(k)$ is represented by
\begin{equation}
Q_2(k)=b_2{\cal I}_2(k)+\frac{1+\delta_c b_1}{\delta_c^2}
\frac{\partial{\cal I}_2(k)}{\partial\log\sigma_0}\;,
\end{equation}
where 
\begin{equation}
{\cal I}_2(k) = \int\!\!\frac{d^3\mathbi{q}}{(2\pi)^3}\, W(qR_s) W(|\mathbi{k}-\mathbi{q}|R_s)
B_s(q,k,|\mathbi{k}-\mathbi{q}|) \;.
\end{equation}
Inserting this expression into $Q_2(k)$ and rearranging the terms, the non-Gaussian halo bias
in the large-scale limit becomes
\begin{equation}
\Delta b_1(k) \approx 2\fnl\calM(k)^{-1}\sigma_\alpha^2
\biggl[\delta_c^{-2}\Bigl(\delta_c^2 b_2+2\delta_c b_1+2\Bigr)
+2\delta_c^{-2}\Bigl(1+\delta_c b_1\Bigr)
\biggl(\frac{\partial\log\sigma_\alpha}{\partial\log\sigma_0}-1\biggr)\Biggr] \;.
\end{equation}
In the case of a Press-Schechter multiplicity function, we have the additional simplifications 
$\delta_c^2 b_2+2\delta_c b_1+2=(\delta_c^3/\sigma_0^2) b_1$ and $1+\delta_c b_1=\nu^2$, so that 
the non-Gaussian bias formula derived by \cite{desjacques/etal:2011a},
\begin{equation}
\Delta b_1(k) = 2\fnl\calM(k)^{-1}\frac{\sigma_\alpha^2}{\sigma_0^2}
\biggl[\delta_c b_1+2\biggl(\frac{\partial\log\sigma_\alpha}{\partial\log\sigma_0}-1\biggr)
\biggr]\;,
\label{eq:djs}
\end{equation}
is recovered.
However, as was noted in \cite{matsubara:2012}, this does not occur in cases other than the 
Press-Schechter mass function (for the Sheth-Tormen mass function \cite{sheth/tormen:1999} for 
instance).

Ref. \cite{ferraro/etal:2012} also explored the clustering of thresholded regions, computing 
2-point statistics from a local bias expansion formulated in terms of Hermite polynomials 
(as done in \cite{matsubara:1995} for Gaussian initial conditions). They showed that such
a local bias expansion leads to a non-Gaussian bias consistent with peak-background split
expectations. This suggests that it should be possible to rephrase the calculations of 
\cite{matsubara:2012,ferraro/etal:2012} in the formalism considered in this paper.

\subsection{Connection with excursion set peaks}

To emphasize the connection with the local bias expansion (\ref{eq:newdpk}) formulated for 
excursion set peaks, let us go one step further and perform the derivative of the step function 
with respect to the smoothing scale $R_s$. The number density of thresholded regions on that filtering 
scale becomes
\begin{equation}
n(\mathbi{x},R_s) = -\frac{2\bar{\rho}}{M}\frac{\partial}{\partial R_s}
\theta_H\!\bigl[\delta_s(\mathbi{x})-\delta_c\bigr]
= \frac{2\bar{\rho}}{M}\frac{\mu(\mathbi{x})}{\sigma_0}\delta_D\!\bigl[\nu(\mathbi{x})-\nu_c\bigr] \;.
\label{eq:npsx}
\end{equation}
For a Gaussian density field, the average number density of thresholded regions thus is
\begin{align}
\bar{n}(R_s) &= \frac{2\bar{\rho}}{M}\sigma_0^{-1}\int\!\!d\nu d\mu\,\mu\delta_D\!(\nu-\nu_c)
\,{\cal N}(\nu,\mu) 
= \frac{2\bar{\rho}}{M}\sigma_0^{-1}\int_{-\infty}^{+\infty}\!\!d\mu\,\mu {\cal N}(\nu_c,\mu) 
= -2 V^{-1}\frac{\sigma_0'}{\sigma_0}\nu_c {\cal N}(\nu_c) \;,
\end{align}
where we have used Bayes' theorem ${\cal N}(\nu_c,\mu)={\cal N}(\mu|\nu_c){\cal N}(\nu_c)$
and the conditional average $\left\la\mu|\nu_c\right\ra\equiv \gamma_{\nu\mu}\nu_c=-\sigma_0'\nu_c$. 
Since $d\nu_c/dR_s=-\nu_c(\sigma_0'/\sigma_0)$, the average number density of thresholded regions 
in the infinitesimal range $[\nu_c,\nu_c+d\nu_c]$ is
\begin{equation}
\bar{n}(\nu_c)=\bar{n}(R_s)\frac{dR_s}{d\nu_c}=-2 V^{-1}\frac{1}{\nu_c\sigma_0'}
\int_{-\infty}^{+\infty}\!\!d\mu\,\mu{\cal N}(\nu_c,\mu)=2 V^{-1}{\cal N}(\nu_c)\;.
\end{equation}
Hence, we get the Press-Schechter multiplicity function 
$f_{\rm PS}(\nu_c)=V\bar{n}_{\rm PS}(\nu_c)=2{\cal N}(\nu_c)$.
Note that the trajectories $\delta(R_s)$ can equally cross the threshold $\delta_c$ up or down
depending on the sign of $\mu$. This leads to the so called ``could-in-cloud'' problem. In our
excursion set peaks approach, this issue is taken care of upon requiring $\mu>0$ [hence the 
multiplicative factor of $\theta_H(\mu)$ in (\ref{eq:nesp})], which turns out to be a very good
approximation at large smoothing scales (where trajectories are nearly fully correlated, see
\cite{musso/paranjape:2012}).

We can now easily convince ourselves that clustering statistics of regions at the threshold 
$\delta_c$ can be computed from the local series expansion
\begin{equation}
\delta_\nu(\mathbi{x}) = b_{10}\delta(\mathbi{x}) + b_{01}\mu(\mathbi{x}) 
+ \frac{1}{2} b_{20}\delta^2(\mathbi{x})
+b_{11}\delta(\mathbi{x})\mu(\mathbi{x})+\frac{1}{2}b_{02}\mu^2(\mathbi{x}) + \cdots 
\label{eq:newdnu}
\end{equation}
with the understanding that, in analogy with discrete peaks, terms involving zero-lag moments 
must be discarded. The bias factors are computed analogously to those of discrete density peaks.
Since $\nu$ and $\mu$ are normally distributed random variables, the bias parameters $b_{ij}$
are bivariate Hermite polynomials averaged over all possible thresholded regions, i.e.
\begin{equation}
\sigma_0^i b_{ij} = 
\frac{1}{\bar{n}(R_s)}\int\!\!d\nu d\mu\,n(\mathbi{x},R_s)H_{ij}(\nu,\mu) {\cal N}(\nu,\mu) \;.
\label{eq:bijdnu}
\end{equation}
Following \cite{desjacques:2012}, the biases $b_{ij}$ can be evaluated straightforwardly from 
the series expansion
\begin{equation}
\Bigl\la f(\epsilon_1,\epsilon_2)\, e^{\epsilon_1\sigma_0 b_\nu +\epsilon_2 b_\mu}
\Bigr\lvert \nu(\mathbi{x})=\nu_c\Bigr\ra 
= \sum_{i,j=0}^\infty \sigma_0^i b_{ij}
\biggl(\frac{\epsilon_1^i}{i!}\biggr)
\biggl(\frac{\epsilon_2^j}{j!}\biggr)\;,
\end{equation}
where $f(\epsilon_1,\epsilon_2)$ is the exponential factor in the bivariate normal 
${\cal N}(\nu,\mu)$ with the replacement $\nu\to\epsilon_1$ and $\mu\to\epsilon_2$, and
\begin{equation}
b_\nu=\frac{1}{\sigma_0}\left(\frac{\Delta_0^2\nu_c+\sigma_0'\mu}{\Delta_0^2-\sigma_0^{'2}}\right)\,,
\qquad
b_\mu=\frac{\mu+\sigma_0'\nu_c}{\Delta_0^2-\sigma_0^{'2}} \;.
\end{equation}
The difference with $b_\nu$ and $b_u$ defined in (\ref{eq:bvbu}) arises from the fact that 
the rms variance of $\mu$ is not normalized to unity. The first-order bias factors thus are
\begin{align}
b_{10} &= -2V^{-1}\Bigl(\nu_c\sigma_0'\bar{n}(\nu_c)\Bigr)^{-1}
\int_{-\infty}^{+\infty}\!\!d\mu\,\frac{\mu}{\sigma_0}
\left(\frac{\Delta_0^2\nu_c+\sigma_0'\mu}{\Delta_0^2-\sigma_0^{'2}}\right){\cal N}(\nu_c,\mu) 
\nonumber \\
&= \frac{1}{\sigma_0}\left(\Delta_0^2-\sigma_0^{'2}\right)^{-1}
\biggl(\Delta_0^2\nu_c-\frac{1}{\nu_c}\left\la\mu^2|\nu_c\right\ra\biggr) \nonumber \\
&= \frac{1}{\sigma_0}\left(\nu_c-\frac{1}{\nu_c}\right) 
\end{align}
for the density $\delta(\mathbi{x})$, and
\begin{align}
b_{01} &=
 -2V^{-1}\Bigl(\nu_c\sigma_0'\bar{n}(\nu_c)\Bigr)^{-1}
\int_{-\infty}^{+\infty}\!\!d\mu\,\frac{\mu}{\sigma_0}
\left(\frac{\mu+\sigma_0'\nu_c}{\Delta_0^2-\sigma_0^{'2}}\right){\cal N}(\nu_c,\mu) 
\nonumber \\
&= \left(\Delta_0^2-\sigma_0^{'2}\right)^{-1}
\biggl(-\frac{1}{\nu_c\sigma_0'}\left\la\mu^2|\nu_c\right\ra+\sigma_0'\nu_c\biggr)
\nonumber \\
&= -\frac{1}{\nu_c\sigma_0'}
\end{align}
for its derivative $\mu(\mathbi{x})$ with respect to the filtering scale. To derive these results, 
we took advantage of the fact that $\la\mu^2|\nu_c\ra=\la\mu|\nu_c\ra^2+\la\Delta\mu^2|\nu_c\ra$,
where $\la\Delta\mu^2|\nu_c\ra=\Delta_0^2-\sigma_0^{'2}$ is the variance at a fixed value
of $\nu=\nu_c$. Consequently, the first-order Fourier space bias factor $c_1(k)$ of 
thresholded regions is (adopting the notational convention of \cite{matsubara:2012}) 
\begin{equation}
c_1(k)\delta(\mathbi{k}) = b_{10}\delta_s(\mathbi{k})+b_{01}\mu(\mathbi{k})
= \left[b_1(M) W+\frac{1}{\delta_c}\frac{\partial W}{\partial\log\sigma_0}\right]\delta(\mathbi{k}) \;,
\end{equation}
which agrees with the expression found in \cite{matsubara:2012,ferraro/etal:2012}.
A similar calculation shows that, at second order, $b_{20}=(\nu_c^2-3)/\sigma_0^2$, 
$b_{11}=-1/(\sigma_0\sigma_0')$ and $b_{02}=0$, so that we recover the bias factor 
$c_2(\mathbi{k}_1,\mathbi{k}_2)$, (\ref{eq:c2k1k2}), in the particular case of thresholded regions 
[for which $b_2(M)=(\nu_c^2-3)/\sigma_0^2$]. This shows that clustering statistics of 
thresholded regions can be computed exactly like those of discrete peaks.

As shown in \cite{desjacques:2012} and in the present work, the effective local bias expansion 
vastly simplifies the calculations for discrete density peaks. Still, computations are somewhat 
more intricate than for thresholded regions. Therefore, the question arises as to whether it 
would be possible to modify the ``localized'' number density (\ref{eq:npsx}) so as to obtain, 
e.g. a Sheth-Tormen multiplicity function and the corresponding bias factors, while 
simultaneously accounting for the correct non-Gaussian bias amplitude. A sensible choice is
\begin{equation}
n(\mathbi{x},R_s)=\frac{2\bar{\rho}}{M}\frac{\mu(\mathbi{x})}{\sigma_0}{\cal W}(\nu,\mu)
\delta_D\!\bigl[\nu(\mathbi{x})-\nu_c\bigr]\;,
\label{eq:npsxw}
\end{equation}
where ${\cal W}(\nu,\mu)$ is some weight function to be determined through the calculation. 
Note that it cannot depend on $\nu$ only, otherwise the bias factors computed from 
(\ref{eq:bijdnu}) would be the same as those of thresholded regions. The average number 
density becomes $\bar{n}(\nu_c)=2 V^{-1}g(\nu_c){\cal N}(\nu_c)$, where
\begin{align}
g(\nu_c) &= -\frac{1}{\nu_c\sigma_0'}\int_{-\infty}^{+\infty}\!\!d\mu\,{\cal W}(\nu_c,\mu)
{\cal N}(\mu|\nu_c) 
= {\cal W}(\nu_c)-\frac{{\cal W}'(\nu_c)}{\nu_c\sigma_0'}\left(\Delta_0^2-\sigma_0^{'2}
+\sigma_0^{'2}\nu_c^2\right) +\cdots \,.
\end{align}
The second equality assumes that ${\cal W}(\nu_c,\mu)$ is Taylor expanded around $\mu=0$.
The first-order Gaussian bias factor $b_{10}$ can be derived either from a peak-background 
split,
\begin{align}
b_{10} = -\frac{1}{\sigma_0\bar{n}}\frac{d\bar{n}}{d\nu_c} &= 
\frac{1}{\sigma_0}\left[\nu_c-\frac{1}{\nu_c}-\frac{g'(\nu_c)}{g(\nu_c)}\right]
=\frac{1}{\sigma_0}\left(\nu_c-\frac{1}{\nu_c}\right)
-a_1 \frac{{\cal W}'(\nu_c)}{{\cal W}(\nu_c)}+\cdots \;,
\end{align}
or from an evaluation of the ensemble average of $b_\nu$, 
\begin{align}
b_{10} &= -\frac{1}{\nu_c\sigma_0' g(\nu_c)}\int_{-\infty}^{+\infty}\!\!d\mu \,
\frac{\mu}{\sigma_0}\frac{\Delta_0^2\nu_c+\sigma_0'\mu}{\Delta_0^2-\sigma_0^{'2}}
{\cal W}(\nu_c,\mu){\cal N}(\mu|\nu_c)
= \frac{1}{\sigma_0}\left(\nu_c-\frac{1}{\nu_c}\right)
-a_2 \frac{{\cal W}'(\nu_c)}{{\cal W}(\nu_c)}+\cdots \;.
\end{align}
A detailed calculation yields $a_1\neq a_2$, which suggests that it may be difficult to ensure 
the equality of both expressions except for very specific choices of ${\cal W}$. Furthermore, 
$b_{02}\neq 0$ so that, unlike thresholded regions, the local bias expansion will also 
involves $\mu^2(\mathbi{x})$-terms etc. Therefore, one cannot simply set $f_{\rm PS}\to f_{\rm ST}$
in the local bias expansion (\ref{eq:newdnu}) [or, equivalently, consider only terms like 
$(\partial/\partial M)(W_1\times \cdots\times W_n)$ as done in \cite{matsubara:2012}].
We will not explore this issue any further here. The bottom line is working out a fully 
consistent biasing scheme based on a fitting formulae for the multiplicity function is not 
trivial.

\subsection{Why does standard local bias fail?}

We are now in a position to address the $b_1$ versus $b_2$ issue mentioned in Section~\ref{sec:intro}. 
As we have seen, the peak-background split ansatz is not absolutely necessary to the 
calculation of clustering statistics of discrete density peaks (and thresholded regions).
However, it is an essential ingredient for the following reasons: i) it can dramatically 
simplify the calculations (through the {\it effective} local bias expansion proposed in 
\cite{desjacques:2012}), and ii) it provides two consistency relations that must be satisfied 
by any realistic (Lagrangian) description of halo clustering. Namely,
\begin{itemize}
\item The $k$-independent piece $b_N$ ($=b_{N00}$ for the excursion set peaks) of the Gaussian 
bias factor $c_N(\mathbi{k}_1,\dots,\mathbi{k}_N)$ is equal to a derivative of the halo mass 
function,
\begin{equation}
b_N = \frac{(-1)^N}{\bar{n}_{\rm h}}\frac{d^N\bar{n}_{\rm h}}{d\delta_c^N} \;.
\end{equation}
\item In the low-$k$ limit, the amplitude of the non-Gaussian correction $\Delta c_1(k)$ to the 
linear bias $c_1(k)$ is proportional to (any proxy for the normalisation of $P_\delta$ is 
acceptable)
\begin{equation}
\Delta c_1(k)\propto \frac{\partial\log\bar{n}_{\rm h}}{\partial\sigma_8} \;.
\end{equation}
\end{itemize}
Even though excursion set peaks and thresholded regions satisfy both equalities, it is generally
difficult to satisfy them simultaneously. 
For example, changing the number density (\ref{eq:npsx}) into (\ref{eq:npsxw}) in an attempt to 
improve the agreement with N-body multiplicity functions appears to violate both conditions. 
This suggests that {\it only Lagrangian bias schemes defined through a set of constraints 
imposed on the linear density field may satisfy these peak-background split relations}.

Therefore, the fundamental reason why the standard Lagrangian local bias model
$\delta_{\rm h}(\mathbi{x})=b_1\delta(\mathbi{x})+b_2\delta^2(\mathbi{x})/2+\cdots$ fails at 
reproducing the correct non-Gaussian amplitude originates from the fact that, unlike the 
discrete density peaks and the thresholded regions considered here, the local bias expansion
is not self-consistently computed from a constrained subset of the linear density field [this 
statement holds true also for the approach of \cite{giannantonio/porciani:2010} which, for 
$\fnl=0$, reduces to standard local bias]. 
The constraints must involve $\delta(\mathbi{x})$ solely for a local bias expansion 
$\delta_{\rm h}(\mathbi{x})={\cal F}[\delta(\mathbi{x})]$ to give results consistent with the 
peak-background split expectations. This is a very restrictive condition given that, e.g. terms 
involving powers of $\mu(\mathbi{x})$ will appear in the series expansion as soon as any barrier 
crossing condition is imposed. 
Since the Gaussian bias $b_1=(\nu_c^2-1)/\delta_c$ of thresholded regions does not agree with 
measurements of the linear halo bias for realistic values of $\nu_c$, we are led to the
inescapable conclusion that a fully consistent modeling of halo clustering statistics with 
Gaussian and non-Gaussian initial conditions requires variables in addition to 
$\delta(\mathbi{x})$ and $\mu(\mathbi{x})$. 
In the case of excursion set peaks, clustering statistics involve also the curvature of the 
density field $u(\mathbi{x})$ etc. (note that it should possible to include such dependencies in 
the approach of \cite{schmidt/etal:2012}). 
The multiplicity function $\fesp(\nu_c)\propto \exp(-\nu_c^2/2)/\nu_c$ guarantees that the 
peak-background split consistency relations hold for both the Gaussian bias parameters and the 
non-Gaussian corrections.
Finally, Lagrangian tidal shear terms may also be significant at low mass (see e.g. 
\cite{sheth/etal:2013} for a quantitative analysis of their influence).

\section{Conclusions}
\label{conclusions}

We have computed non-Gaussian corrections arising from the primordial bispectrum to the 2-point
correlation function of discrete density peaks. We have shown that the local bias expansion
of \cite{desjacques:2012} gives rise to the same result as a direct calculation from the 
Edgeworth expansion. Furthermore, we have generalized the former to the excursion set peaks of 
\cite{paranjape/sheth:2012}. 
In all cases, we have checked that the non-Gaussian correction to the linear peak bias agrees 
with peak-background split expectations. 
More precisely, the effective local bias expansion predicts a non-Gaussian amplitude which is 
a sum over quadratic bias parameters. All these terms nicely combine into a derivative of 
$\log\bar{n}_{\rm h}$ with respect to the normalisation amplitude $\sigma_8$ (or, equivalently, 
the halo mass $M$).
This is very encouraging since excursion set peaks have recently been shown to reproduce very 
well the Gaussian mass function and bias of dark matter halos 
\cite{paranjape/sheth/desjacques:2012}. 
Even though we have focused on the contribution of the primordial bispectrum, our analysis can 
be straightforwardly extended to generic non-Gaussian initial conditions, and provide the
basis for a computation of non-Gaussian corrections to the halo bispectrum.

Our results shed new light on the widespread local Lagrangian bias expansions and their 
ability to model Gaussian and non-Gaussian halo clustering statistics. 
We argue that generic Lagrangian bias expansions cannot properly model these statistics (and, 
thereby, satisfy the peak-background split relations mentioned above) unless the series expansion 
consistently arises from a set of constraints applied to the linear density field (as is the 
case for peaks or thresholded regions). 
Since the trivial constraint $\delta_s(\mathbi{x})>\delta_c$ does not furnish a good fit to halo 
statistics, we are inescapably led to consider additional variables beyond the density $\delta_s$ 
and, therefore, local Lagrangian bias relations beyond the widespread series expansion 
$\delta_{\rm h}(\mathbi{x})={\cal F}[\delta_s(\mathbi{x})]$.

\acknowledgments

We would like to thank Donghui Jeong, Fabian Schmidt, Roman Scoccimarro, Ravi Sheth and Shuichiro 
Yokoyama for many stimulating discussions on bias and primordial non-Gaussianity. 
VD acknowledges support by the Swiss National Science Foundation.
JG was supported by a Korean-CERN fellowship while this work was under progress, and acknowledges 
the Max-Planck-Gesellschaft, the Korea Ministry of Education, Science and Technology, 
Gyeongsangbuk-Do and Pohang City for the support of the Independent Junior Research Group at the 
Asia Pacific Center for Theoretical Physics.
AR is supported by the Swiss National Science Foundation (SNSF), project ``The non-Gaussian 
Universe'' (project number: 200021140236).

\bibliographystyle{JHEP}
\bibliography{references}

\end{document}